\documentclass[11pt]{article}
\usepackage{xparse}
\usepackage{amsmath}
\usepackage{amssymb}
\usepackage{textcomp}
\usepackage{ragged2e}
\usepackage{setspace}
\usepackage{comment}
\usepackage{slashed}
\usepackage{etoolbox}
\usepackage[affil-it]{authblk}
\usepackage{ifthen}

\usepackage[pdfencoding=auto, psdextra]{hyperref}
 \hypersetup{
     colorlinks=true,
     linkcolor=blue,
     citecolor = blue,      
     urlcolor=blue,
     }
\newenvironment{changemargin}[2]{
\begin{list}{}{
\setlength{\topsep}{0pt}
\setlength{\leftmargin}{#1}
\setlength{\rightmargin}{#2}
\setlength{\listparindent}{\parindent}
\setlength{\itemindent}{\parindent}
\setlength{\parsep}{\parskip}
}
\item[]}{\end{list}}

\newcommand{\bb}[1]{\mathbb{#1}}
\newcommand{\alg}[1]{\mathfrak{#1}}
\newcommand{\grp}[1]{\operatorname{#1}}
\newcommand{\R}{\bb{R}}
\newcommand{\C}{\bb{C}}
\newcommand{\Hq}{\bb{H}}
\newcommand{\Z}{\bb{Z}}
\newcommand{\M}[2]{\grp{M}_{#1}(#2)}
\newcommand{\K}{\kappa}

\newcommand{\GL}[2][]{%
   \ifthenelse{ \equal{#1}{} }
      {\grp{GL}({#2})}
      {\grp{GL}^{#1}({#2})}
}
\newcommand{\U}[2][]{%
   \ifthenelse{ \equal{#1}{} }
      {\grp{U}({#2})}
      {\grp{U}^{#1}({#2})}
}
\newcommand{\uu}[2][]{%
   \ifthenelse{ \equal{#1}{} }
      {\alg{u}({#2})}
      {\alg{u}^{#1}({#2})}
}
\newcommand{\SL}[2][]{%
   \ifthenelse{ \equal{#1}{} }
      {\grp{SL}({#2})}
      {\grp{SL}^{#1}({#2})}
}
\newcommand{\SO}[2][]{%
   \ifthenelse{ \equal{#1}{} }
      {\grp{SO}({#2})}
      {\grp{SO}^{#1}({#2})}
}
\newcommand{\SU}[2][]{%
   \ifthenelse{ \equal{#1}{} }
      {\grp{SU}({#2})}
      {\grp{SU}^{#1}({#2})}
}
\newcommand{\Spin}[2][]{%
   \ifthenelse{ \equal{#1}{} }
      {\grp{Spin}({#2})}
      {\grp{Spin}^{#1}({#2})}
}
\newcommand{\Pin}[2][]{%
   \ifthenelse{ \equal{#1}{} }
      {\grp{Pin}({#2})}
      {\grp{Pin}^{#1}({#2})}
}
\newcommand{\Orth}[2][]{%
   \ifthenelse{ \equal{#1}{} }
      {\grp{O}({#2})}
      {\grp{O}^{#1}({#2})}
}
\newcommand{\PO}[2][]{%
   \ifthenelse{ \equal{#1}{} }
      {\grp{PO}({#2})}
      {\grp{PO}^{#1}({#2})}
}
\newcommand{\PSO}[2][]{%
   \ifthenelse{ \equal{#1}{} }
      {\grp{PSO}({#2})}
      {\grp{PSO}^{#1}({#2})}
}
\newcommand{\SP}[2][]{%
   \ifthenelse{ \equal{#1}{} }
      {\grp{SP}({#2})}
      {\grp{SP}^{#1}({#2})}
}
\newcommand{\Sp}[2][]{%
   \ifthenelse{ \equal{#1}{} }
      {\grp{Sp}({#2})}
      {\grp{Sp}^{#1}({#2})}
}
\newcommand{\SK}[2][]{%
   \ifthenelse{ \equal{#1}{} }
      {\grp{SO}^*({#2})}
      {\grp{SO}^*^{#1}({#2})}
}
\newcommand{\cl}[2][]{%
   \ifthenelse{ \equal{#1}{} }
      {\grp{C\ell}({#2})}
      {\grp{C\ell}^{#1}({#2})}
}
\newcommand{\sk}[2][]{%
   \ifthenelse{ \equal{#1}{} }
      {\alg{so}^*({#2})}
      {\alg{so}^*^{#1}({#2})}
}
\newcommand{\symp}[2][]{%
   \ifthenelse{ \equal{#1}{} }
      {\alg{sp}({#2})}
      {\alg{sp}^{#1}({#2})}
}
\newcommand{\spl}[2][]{%
   \ifthenelse{ \equal{#1}{} }
      {\alg{sl}({#2})}
      {\alg{sl}^{#1}({#2})}
}
\newcommand{\so}[2][]{%
   \ifthenelse{ \equal{#1}{} }
      {\alg{so}({#2})}
      {\alg{so}^{#1}({#2})}
}
\newcommand{\su}[2][]{%
   \ifthenelse{ \equal{#1}{} }
      {\alg{su}({#2})}
      {\alg{su}^{#1}({#2})}
}
\newcommand{\spin}[2][]{%
   \ifthenelse{ \equal{#1}{} }
      {\alg{spin}({#2})}
      {\alg{spin}^{#1}({#2})}
}

\title{The Anti-Unitarity of Time Reversal \& Co-representations of Lorentzian Pin Groups}
\author{Craig M$^{\mathrm{c}}$Rae\thanks{Electronic address: \texttt{mcraec3@myumanitoba.ca}}}
\affil{University of Manitoba \\ Winnipeg, MB}

\begin{document}
\maketitle
  \pagenumbering{gobble}
  \pagenumbering{arabic} 
\setlength{\parindent}{1.5em}
\setlength{\parskip}{0.5ex}
\begin{changemargin}{-1.5cm}{-1.5cm}

\begin{abstract}
\noindent \normalsize In the representation theory of Lorentzian orthogonal groups, there are well known arguments as to why the parity inversion operator $\mathcal{P}$ and the time reversal operator $\mathcal{T}$, should be realized as linear and anti-linear operators respectively (Wigner 1932). Despite this, standard constructions of double covers of the Lorentzian orthogonal groups naturally build time reversal operators in such a manner that they are linear, and the anti-linearity is put in ad-hoc after the fact. This article introduces a viewpoint naturally incorporating the anti-linearity into the construction of these double covers, through what Wigner called co-representations, a kind of semi-linear representation. It is shown how the standard spinoral double covers of the Lorentz group --- $\operatorname{Pin}(1,3)$ and $\operatorname{Pin}(3,1)$ --- may be naturally centrally extended for this purpose, and the relationship between the $\mathcal{C}$, $\mathcal{P}$, and $\mathcal{T}$ operators is discussed. Additionally a mapping is constructed demonstrating an interesting equivalence between Majorana and Weyl spinors. Finally a co-representation is built for the de Sitter group $\Pin{1,4}$, and it is demonstrated how a theory with this symmetry has no truly scalar fermion mass terms.
\end{abstract}
\end{changemargin}
\begin{changemargin}{-2cm}{-2cm}


\newpage 
\tableofcontents

\newpage

\baselineskip=1.2\baselineskip
\section*{Introduction}
\label{sec:Preamble}
\addcontentsline{toc}{section}{\nameref{sec:Preamble}}
Pin groups are to spin groups, as orthogonal groups are to special orthogonal groups. Where spin groups allow us to understand how to perform rotations and boosts upon spinors, pin groups allow us to perform reflections upon spinors. When behaviour under reflections is being emphasized, one may refer to these as \textit{pinors} instead.\footnote{It is worthwhile to point out that as the pin groups are `larger' than the spin groups, the irreducible representations of pin groups, upon restriction to the spin subgroup, will generically be reducible spin representations; pin irreps and spin irreps are not one-to-one.} It is interesting, though perhaps inconvenient, that there are two non-isomorphic pin groups for any one spin group; these correspond to the two choices of metric signature for any orthogonal group $\Orth{p,q} \equiv \Orth{q,p}$, and are called $\Pin{p,q}$ and $\Pin{q,p}$. In particular below we will explore the structure of the groups $\Pin{3,1}$ and $\Pin{1,3}$, both serving as double coverings of the entire Lorentz group $\Orth{1,3}$. We will see these groups are not the whole story, as these constructions unfortunately build all reflections in a necessarily linear way, violating certain demands of any reasonable time reversal operator as outlined in \cite{WIGNERBOOK}. In particular one needs an apparatus for constructing double covers of orthogonal groups where some elements act anti-linearly. In Sec.~(\ref{sec:CPTcorep}) we will see how to extend the pin groups as naturally as possible in order to access anti-linear operations, related to the introduction of an intertwining $\mathcal{CPT}$ operator.

As in the classical case \cite{ConvLorRep}, we have the intent to view the pin groups as a semi-direct product between the identity component of the spin group and a discrete reflection group. The different signatures will correspond to different discrete reflection groups of the same order. Thus once the relevant spin group is understood there are principally only three reflections whose behaviours need to be understood in order to build the relevant pin groups: spacial inversion ($\mathcal{P}$), time reversal ($\mathcal{T}$), and their product ($\mathcal{PT}$). Importantly for these constructions, the pin groups are defined as particular groups within a Clifford algebra, utilized below to build and contrast the two pin groups of interest (for a great introduction to the study of Clifford algebras and their relationship to spinors and orthogonal groups see Vaz \& Rocha \cite{CliffAlg}). It is assumed the reader is familiar with Clifford algebra's, and with the representation theory of $\Orth{1,3}$ as well as the double cover of its identity component, $\Spin[+]{1,3}$.

\newpage
\section{The Pin Groups}
Pin groups can be constructed for any signature $\Orth{p,q}$ by first building the relevant Clifford algebra. This is the formal construction which generalizes the Pauli matrices and Dirac matrices. For some metric components $\eta^{\mu\nu}$ with signature $(p,q)$, the Clifford algebra is the smallest real matrix algebra generated by elements $\gamma^\mu$ satisfying the anti-commutation relations:
\begin{equation}
    \gamma^\mu \gamma^\nu + \gamma^\nu \gamma^\mu = 2 \eta^{\mu\nu}.
\end{equation}
For our two cases of interest we have the Minkowski metric $\eta^{\mu\nu} = \pm\operatorname{diag}\{-1,1,1,1\}$. The algebra is spanned not only by sums of $\gamma^\mu$ but also all products thereof, which gives a total real dimension of $2^n$, and it is known that every Clifford algebra is isomorphic to matrix algebra over the reals, complex numbers, or quaternions. Importantly every Clifford algebra is $\mathbb{Z}_2$ graded: its elements may be split into even and odd parts. Of this decomposition the even elements form a subalgebra $\cl[0]{p,q}$: those elements which are spanned only by even numbers of products of the $\gamma^\mu$. In a standard basis the pin group may be taken as the subgroup of invertible Clifford algebra elements which have a determinant of $\pm1$. The spin group is the same, but restricted to the even subalgebra. We will now construct these below for the two signatures.

\subsection{\texorpdfstring{$\Pin{3,1}$}{Pin(3,1)}}
In this case we take $\eta^{\mu\nu} = \operatorname{diag}\{-1,1,1,1\}$. From the classification of Clifford algebras we know that $\cl{3,1} \cong \M{4}{\R}$: the matrix algebra of real $4\times4$ matrices, which is $16$ dimensional. One choice of basis for the $\gamma^\mu$ is given below:
\begin{equation}
\label{eq:CL31basis}
\begin{split}
\quad \gamma^0 &=
    \begin{pmatrix}
        0 & 0 & 0 & -1 \\
        0 & 0 & -1 & 0 \\
        0 & 1 & 0 & 0 \\
        1 & 0 & 0 & 0 
    \end{pmatrix}, \quad \gamma^1 = 
        \begin{pmatrix}
        0 & 1 & 0 & 0 \\
        1 & 0 & 0 & 0 \\
        0 & 0 & 0 & -1 \\
        0 & 0 & -1 & 0 
    \end{pmatrix}, \quad \gamma^2 =    \begin{pmatrix}
        1 & 0 & 0 & 0 \\
        0 & -1 & 0 & 0 \\
        0 & 0 & 1 & 0 \\
        0 & 0 & 0 & -1 
    \end{pmatrix}, \\ \gamma^3 &=    \begin{pmatrix}
        0 & 0 & 0 & -1 \\
        0 & 0 & -1 & 0 \\
        0 & -1 & 0 & 0 \\
        -1 & 0 & 0 & 0 
    \end{pmatrix} ,\quad 
     \gamma^5 = \gamma^0\gamma^1\gamma^2\gamma^3 = \begin{pmatrix}
        0 & -1 & 0 & 0 \\
        1 & 0 & 0 & 0 \\
        0 & 0 & 0 & -1 \\
        0 & 0 & 1 & 0 
    \end{pmatrix}.
\end{split}    
\end{equation}
The first four gamma matrices generate the entire Clifford algebra: the span of these matrices and their products is $\M{4}{\R}$. The fifth gamma matrix is given for convenience, though it should be noted in the interest of explicitly preserving real structures, the conventional (physics literature) factor of $i$ is not included in $\gamma^5$, and so $(\gamma^5)^2 = -1$. 

Before the introduction of gauge charges, the \textit{pinor} states these matrices act upon may be taken to be totally real, with only four real components, exactly as with Majorana spinors, and we will see later in Sec.~(\ref{sec:majasweyl}) there is an interesting relationship between these pinors and the representation of Weyl spinors. Let us now build the spin group and then determine the nature of the $\mathbf{P}$ and $\mathbf{T}$ operators, to have a better grasp of these pinors.

\subsubsection{\texorpdfstring{$\Spin[+]{3,1} \subset \cl[0]{3,1}$}{Spin+(3,1) in CL(3,1)}}
\label{sec:spin31}
One of the main conveniences offered by the Clifford algebra construction is that it simultaneously contains representations of the Lie algebra $\spin{3,1}$ as well as the corresponding group $\Spin[+]{3,1}$: these are both contained within the even subalgebra; the exponential map between them is defined in the algebra via its power series. The generators $s^{\mu\nu}$ of $\spin{3,1}$, with $\gamma^\mu$ understood above as the standard Cartesian basis vectors in Minkowski spacetime, are given via:\footnote{It is interesting that the canonical convention for right handed generators is `unnatural' in this real Clifford algebra, hence the spurious minus sign.}
\begin{equation}
    s^{\mu\nu} = -\frac{1}{4} \left[\gamma^\mu, \gamma^\nu\right] = -\frac{1}{2} \gamma^\mu \gamma^\nu.
\end{equation}
In particular, the Cartesian generators of rotation $L_i$ and boosts $K_i$ are:
\begin{equation}
    L_i = - \frac{1}{2} \varepsilon_{ijk}\gamma^j \gamma^k, \qquad K_i = -\frac{1}{2}\gamma^0 \gamma^i.
\end{equation}
These six generators span the Lie algebra $\spin{3,1}$, for which products of exponents thereof will give every element of $\Spin[+]{3,1}$. It is not difficult to verify this representation is a double covering of $\SO[+]{3,1}$ as rotations by $2\pi$ give $-1$. Thus we have the identity component of $\Pin{3,1}$ figured out. The next easiest island of transformations to find, is the double covering of $\SO[-]{p,q}$, done by looking at the center of these groups.

\subsubsection{\texorpdfstring{$\Spin{3,1} \subset \cl[0]{3,1}$}{Spin(3,1) in CL(3,1)}}
Within the Clifford algebra there is a non-trivial element which commutes with every generator of $\spin{3,1}$, and thus commutes with the whole group $\Spin[+]{3,1}$: that element is $\gamma^5$. The naming convention is deceptive since $\gamma^5 = \gamma^0\gamma^1\gamma^2\gamma^3$ is an even element of the Clifford algebra; yet simple orthogonality arguments show it is not contained within $\Spin[+]{3,1}$. As it is of unit determinant and within the even subalgebra, it must be a member of the larger group $\Spin{3,1}$: the double cover of \textit{both} disconnected components of $\SO{3,1}$. By definition the elements of $\Spin{3,1}$ form a group homomorphism to $\SO{3,1}$, and so the center of $\Spin{3,1}$ is mapped into the center of $\SO{3,1}$. The trivial element of the center of $\SO[+]{3,1}$, the identity transform, is already double covered by $\pm1 \in \Spin[+]{3,1}$, and as $\pm \gamma^5$ is central in $\Spin{3,1}$, it must be mapped to $-1 = \mathcal{PT} \in \SO{3,1}$ under the double covering. Formally the structure of the groups is:
\begin{equation}
\label{eq:fullSpin}
    \Spin{3,1} = \left(\Spin[+]{3,1} \times \Z_4\right)/\Z_2.
\end{equation}
This says that to build the full spin group, start with the identity component, and then allow multiplication by $\gamma^5$ (this is the generator of the $\Z_4$); after this modulo out by a $\Z_2$, formally demanding that the $-1$ which $\gamma^5$ squares to, is the same $-1$ which is inside $\Spin[+]{3,1}$. Notice this argument did not depend on the signature at hand, only the commutation relations.

It is worth pointing out that if all we were interested in was the connected component Spin$^+$, as physicists often are, the fact that $\gamma^5$ commutes with every element but is not a multiple of the identity, indicates that over a \textit{complex} vector space this construction is a reducible representation: the well known left-right decomposition. However the construction is irreducible if we restrict to only real vector spaces. Regardless we are interested in the whole pin group, so next we must find the parity and time reversal operators.

\subsubsection{\texorpdfstring{From $\Spin{3,1}$ to $\Pin{3,1}$}{From Spin(3,1) to Pin(3,1)}}
Typical interpretation of Clifford algebra elements would imply $\gamma^1\gamma^2\gamma^3$ should be a spacial inversion, and $\gamma^0$ should be a time reversal. However we will see the Lorentz group has somewhat more freedom in this choice than is typical. The Lie algebra of the Lorentz group is enough to narrow down the possible operators to those which could be either $\mathbf{P}$ or $\mathbf{T}$; then we may use our knowledge of $\mathbf{PT}$ above to constrain closer to a definite answer. Physically, the action of parity or time reversal (but not both) should commute with generators of rotations, and invert (anti-commute with) generators of boosts:
\begin{equation}
\mathbf{P} L_i = L_i \mathbf{P}, \quad \mathbf{P} K_i =  -K_i \mathbf{P}, \qquad \mathbf{T} L_i = L_i \mathbf{T}, \quad \mathbf{T} K_i =  -K_i \mathbf{T}.
\end{equation}
In our search for a pin representation of $\mathbf{P}$ and $\mathbf{T}$, we are looking for elements of the Clifford algebra $\cl{3,1}$ with this behaviour. Commutation with rotation generators $L_i$ means the element must contain either none or both relevant gamma matrices of all three generators. This restricts the possibilities to combinations (sums and products) of $\gamma^0$ and $\gamma^1\gamma^2\gamma^3$, or more conveniently thought of as the span of $\gamma^0$, $\gamma^5$, and $\gamma^0 \gamma^5$. The second condition allows us to eliminate $\mathbf{PT} = \gamma^5$ from the above, as it commutes with boosts. Thus our $\mathbf{P}$ and $\mathbf{T}$ operators lay somewhere in the `unit circle' of the span of $\gamma^0$ and $\gamma^0 \gamma^5 = -\gamma^1\gamma^2 \gamma^3$. One way to write them generically is through the parametrization below
\begin{equation}
    \mathbf{P} = \gamma^0 e^{\gamma^5\theta }, \quad \mathbf{T} = \gamma^0 e^{\gamma^5 \phi}.
\end{equation}
For their product $\mathbf{PT}$ to be $\gamma^5$, the angles must be off by $\pi/2$, leading to
\begin{equation}
    \mathbf{P} = \gamma^0 e^{\gamma^5\theta }, \quad \mathbf{T} = \gamma^5  \gamma^0 e^{\gamma^5 \theta }.
\end{equation}
The final restriction requires an arbitrary choice. The aforementioned typical Clifford algebra approach would correspond to $\theta = \pi/2$. Perhaps surprisingly this more natural choice is not the standard one, and nearly all treatments of the topic an alternate convention is taken with $\theta=0$.  The standard assignment gives
\begin{equation}
    \mathbf{P} = \gamma^0, \quad \mathbf{T} = \gamma^5 \gamma^0, \quad \mathbf{PT} = \gamma^5,
\end{equation}
and the more natural approach would simply swap $\mathbf{P}$ and $\mathbf{T}$. The overall sign of any of these operators is arbitrary. Importantly note that each of these `primary reflections' square to $-1$:
\begin{equation}
\label{eq:PTQuat}
    \mathbf{P}^2 = -1, \quad \mathbf{T}^2 = -1, \quad \left(\mathbf{PT}\right)^2 = -1.
\end{equation}
Combined with the fact each of these three mutually anti-commute, we can see that $\mathbf{P}$ and $\mathbf{T}$ generate a representation of the group of unit quaternions, called $\grp{Q}_8$. 

Every element of $\Pin{3,1}$ may be built as the composition of one of these primary reflections, and an element of $\Spin[+]{3,1}$. In particular the structure of the pin group is given formally via
\begin{equation}
    \Pin{3,1} = \left(\Spin[+]{3,1} \rtimes \grp{Q}_8\right)/\Z_2.
\end{equation}
In the semi-direct product, $\pm \mathbf{P}, \>\pm \mathbf{T} $ act on the Lorentz group via the outer automorphism which inverts boosts $K_i \mapsto -K_i$. Of note is that the center of the group $\Pin{3,1}$ is actually smaller than that of $\Spin{3,1}$, because $\mathbf{P}$ and $\mathbf{T}$ anti-commute with $\mathbf{PT}= \gamma^5$ we have `lost' that element of the center by including these reflections. This is not a problem, as the elements of the double covering which are mapped to the center of $\grp{O}(3,1)$ need only commute up to a sign for the covering to be a homomorphism. The center of the spin group is $\mathcal{Z}(\Spin{3,1}) = \Z_4 = \{\pm\gamma^5, \pm 1\}$. The center of the pin group is $\mathcal{Z}(\Pin{3,1}) = \Z_2 = \pm 1$. 

\subsection{\texorpdfstring{$\Pin{1,3}$}{Pin(1,3)}}
Switching signatures now we construct the other pin group. In this case we take $\eta^{\mu\nu} = \operatorname{diag}\{1,-1,-1,-1\}$. From the classification of Clifford algebras one can see that $\cl{1,3} \cong \M{2}{\Hq}$. This Clifford algebra is the \textit{real} matrix algebra of quaternionic $2\times2$ matrices, which is $16$ dimensional. A basis of generators for $\cl{1,3}$, our quaternionic gamma matrices, reusing the same symbols, is given below:
\begin{equation}
\begin{split}
    \gamma^0 = \begin{pmatrix}
        1 & 0 \\
        0 & -1 
    \end{pmatrix}, \quad \begin{Bmatrix}
        \gamma^1 \\
        \gamma^2 \\
        \gamma^3
    \end{Bmatrix} =  \begin{Bmatrix}
        i \\
        j \\
        k
    \end{Bmatrix}\begin{pmatrix}
        0 & 1 \\
        1 & 0 
    \end{pmatrix}, \quad \gamma^5 = \gamma^0\gamma^1\gamma^2\gamma^3 = \begin{pmatrix}
        0 & -1 \\
        1 & 0
    \end{pmatrix}
\end{split}
\end{equation}
These four generators play the same role as those in the above case, generating the entire algebra through products and sums thereof. Notice we now have that $(\gamma^0)^2 = 1$ while $(\gamma^i)^2 = -1$. The volume element $\gamma^5$ defines another anti-commuting matrix, squaring to $-1$ again in this case. Just as above we may define the $\spin{1,3}$ algebra:
\begin{equation}
    s^{\mu\nu} = \frac{1}{4} \left[\gamma^\mu, \gamma^\nu\right] = \frac{1}{2} \gamma^\mu \gamma^\nu.
\end{equation}
In particular
\begin{equation}
    \begin{Bmatrix}
        L_x \\
        L_y \\ 
        L_z 
    \end{Bmatrix} = \frac{1}{2}     \begin{Bmatrix}
        i \\
        j \\ 
        k 
    \end{Bmatrix} \begin{pmatrix}
        1 & 0 \\ 
        0 & 1
    \end{pmatrix} ,\qquad     \begin{Bmatrix}
        K_x \\
        K_y \\ 
        K_z 
    \end{Bmatrix} = \frac{1}{2}     \begin{Bmatrix}
        i \\
        j \\ 
        k 
    \end{Bmatrix} \begin{pmatrix}
        0 & 1 \\ 
        -1 & 0
    \end{pmatrix}.
\end{equation}
It is easy to verify here this is a projective representation of $\SO[+]{1,3}$ as the commutation relations are the standard ones while rotations by $2 \pi$ give $-1$.
\subsubsection{\texorpdfstring{From $\Spin{1,3}$ to $\Pin{1,3}$}{From spin(1,3) to pin(1,3)}}
Note that the signature does not affect the spin group: $\Spin{3,1} \equiv \Spin{1,3}$. Thus the arguments from Sec.~(\ref{sec:spin31}) are ambivalent to the signature at play, relying only the commutation relations to make conclusions, leaving us identical restrictions and choices from a symbolic point of view, so once again we must have 
\begin{equation}
    \mathbf{PT} = \gamma^5
\end{equation}
and then once again we will choose
\begin{equation}
    \mathbf{P} = \gamma^0, \quad \mathbf{T} = \gamma^0\gamma^5.
\end{equation}
However notice this time the discrete reflection group is different with these different gamma matrices:
\begin{equation}
\label{eq:PTDih}
    \mathbf{P}^2 = 1, \quad \mathbf{T}^2 = 1, \quad \left(\mathbf{PT}\right)^2 = -1.
\end{equation}
The behaviour of the discrete group generated by $\mathbf{P},\mathbf{T},$ and $\mathbf{PT}$ form the structure of $\grp{D}_8$, the dihedral group of order $8$. Initially this construction is concerning, because a time reversal operator should not have possible eigenvectors in spin representations (since spin is always odd under time reversal), and so the square is physically wrong; this problem will go away when incorporating anti-linearity. Every element of $\Pin{1,3}$ may be written as one of these primary reflections, followed by an element of $\Spin[+]{1,3}$, giving the pin group as a semi-direct product
\begin{equation}
    \Pin{1,3} = \left(\Spin[+]{3,1} \rtimes \grp{D}_8\right)/\Z_2.
\end{equation}
Of course the product is semi-direct because $\pm\mathbf{P}$ and $\pm\mathbf{T}$ have the effect of the outer automorphism on the connected component, inverting boosts.

\section{The Discrete Reflection Groups: \texorpdfstring{$\grp{Q}_8$, $\grp{D_8}$, and $\grp{K}_4$}{Q8, D8, and K4}}
\begin{table}[h!]
    \centering
    \begin{tabular}{|c|c|c|}
    \hline Double Cover of $\grp{K}_4$ & Abelian? & \# of Elements which square to $-1$\\\hline
       $\Z_2 \times\Z_2 \times\Z_2$  &  Yes & 0 \\
        $\Z_2\times \Z_4$ & Yes & $4$ \\
        $\grp{D}_8$ & No & $2$\\
        $\grp{Q}_8$ & No & $6$\\\hline 
    \end{tabular}
    \caption{The double covers of the Klein group. In order they are: two copies of the Klein group, two copies of the cyclic group of order four, the Dihedral group of order $8$, and the group of unit quaternions.}\vspace{-0.5em}
    \label{tab:K4doublecovers}
\end{table}
\noindent The quaternion group $\grp{Q}_8$ and the dihedral group $\grp{D_8}$ are siblings. They are two of the five groups of order $8$, and the only of those $5$ which are non-Abelian. Importantly they are both double coverings of the Klein four group $\grp{K}_4$. Taking either of these groups and moduloing by a $\Z_2$, identifying elements related by application of $-1$, one finds the Klein group. The pin groups not only contain double covers of the special orthogonal group, but also of the discrete reflection group. This close relationship between $\grp{Q}_8$ and $\grp{D}_8$ results in them having essentially the same representation theory. Both groups have exactly $5$ irreducible representations over $\mathbb{C}$: four one dimensional representations merely `extending' the representations of $\grp{K}_4$, and a faithful two dimensional representation under which pinors are transformed.

\newpage
\subsection{\texorpdfstring{$\grp{Q}_8$ Representation Theory for $\Pin{3,1}$}{Q8 Representation Theory for Pin(3,1)}}
\begin{table}[h]
    \hspace{-4em}\Centering
    \begin{tabular}{|c|cccc|}\hline
        Name & $\pm 1$ & $\pm \mathbf{P}$ & $\pm \mathbf{T}$ & $\pm \mathbf{PT}$ \\\hline
        $\rho_1$ & $1$ & $1$ & $1$ & $1$ \\
        $\rho_T$ & $1$ & $1$ & $-1$ & $-1$ \\
        $\rho_P$ & $1$ & $-1$ & $1$ & $-1$ \\
        $\rho_{PT}$ & $1$ & $-1$ & $-1$ & $1$ \\ \hline
        $\rho_S$ & $\pm\begin{pmatrix}
            1 & 0 & 0 & 0 \\
            0 & 1 & 0 & 0 \\
            0 & 0 & 1 & 0 \\
            0 & 0 & 0 & 1 \\
        \end{pmatrix}$ & $\pm\begin{pmatrix}
            0 & 0 & 0 & 1 \\
            0 & 0 & 1 & 0 \\
            0 & -1 & 0 & 0 \\
            -1 & 0 & 0 & 0 \\
        \end{pmatrix}$ & $\pm\begin{pmatrix}
            0 & 0 & -1 & 0 \\
            0 & 0 & 0 & 1 \\
            1 & 0 & 0 & 0 \\
            0 & -1 & 0 & 0 \\
        \end{pmatrix}$ & $\pm\begin{pmatrix}
            0 & -1 & 0 & 0 \\
            1 & 0 & 0 & 0 \\
            0 & 0 & 0 & -1 \\
            0 & 0 & 1 & 0 \\
        \end{pmatrix}$ \\\hline
    \end{tabular}
    \caption{Irreducible Representations of $\grp{Q}_8$ over $\R$. The one dimensional representations are labeled by which elements act as $-1$, and $\rho_S$ the `spinor' rep, is the largest and only possible real faithful irrep.}
    \label{tab:Q8reps}
\end{table}

\noindent Despite what was said above, the representations of $\grp{Q}_8$ in Tab.~(\ref{tab:Q8reps}) are given over $\R$ instead of $\C$ because $\Pin{3,1}$ is manifestly real and so the aforementioned two dimensional representation of $\grp{Q}_8$ over the complex numbers is promoted to a four dimensional representation over the reals. This representation will be indicated by an $S$, for spinor. The one dimensional representations are simply $\Z_2$ extensions of the representations of $\grp{K}_4$, and the `interesting' representation, the spinor representation $\rho_S$, is given in the basis of $\Pin{3,1}$ above. As the tensor products of the one dimensional representations are the same as the Klein group, the only knowledge to fill in is how the tensor products of representations interact with $\rho_S$. Another way to ask this, is if one has two pinors $\psi$, $\phi$ upon which $\rho_S$ acts, what is the behaviour of $\psi \otimes \phi$ under these reflections? 

From the Clifford algebra we know the tensor product may be decomposed into 16 components:\footnote{This decomposition is not as pretty as one might expect, and this is because we have taken merely the transpose in the outer product, instead of one of the standard adjoints of the spinor, which we will do soon (and see why it was avoided at first). As well, slightly nonstandard expressions of basis gamma matrices are here to keep the index notation free of any Levi-Civita symbols so we may avoid confusion when the standard decomposition is recovered.}
\begin{equation}
\label{eq:genericdecomp}
    \psi \phi^\intercal = s \gamma^0 + p_0 \>\mathbb{I}_4 + p_{i} \gamma^0 \gamma^i + G_{0i} \gamma^i  +G_{ij} \gamma^i\gamma^j \gamma^0 + v_0 \gamma^5+ v_{i} \gamma^i\gamma^5 \gamma^0 + a \gamma^0 \gamma^5,
\end{equation}
where $s, p_{\mu}, G_{\mu\nu}, v_\mu, a$ are all real coefficients. Importantly the coefficients are for more than convenience: the components $p_\mu$ really do transform like a four-vector under rotations and boosts, when rotating and boosting the pinors $\psi, \phi$ via $\Spin[+]{3,1}$. The same for the other scalar, vector, and tensor components. The commutation relations of the gamma matrices will allow us to quickly see how these components are altered by the elements of the discrete reflection group.  In particular it can be seen:
\begin{equation}
\begin{split}
\label{eq:Q8PtensorDecomp}
    \mathbf{P}\psi \phi^\intercal \mathbf{P}^\intercal &= s \gamma^0 + p_0 \>\mathbb{I}_4 - p_{i} \gamma^0 \gamma^i - G_{0i} \gamma^i + G_{ij} \gamma^i\gamma^j \gamma^0 - v_0 \gamma^5 + v_{i} \gamma^i\gamma^5 \gamma^0 - a \gamma^0 \gamma^5,\\
    \mathbf{T}\psi \phi^\intercal \mathbf{T}^\intercal&= -s \gamma^0 + p_0 \>\mathbb{I}_4 - p_{i} \gamma^0 \gamma^i + G_{0i} \gamma^i  -G_{ij} \gamma^i\gamma^j \gamma^0 - v_0 \gamma^5+ v_{i} \gamma^i\gamma^5 \gamma^0 + a \gamma^0 \gamma^5.
\end{split}
\end{equation}
Note that a different sign on $\mathbf{P}$ or $\mathbf{T}$ would simply cancel, and so this tells us all the required information about the $8$ elements of $\grp{Q}_8$ in this representation. The main takeaway here is that the irreducible action upon the spinors, descends to an action upon the tensor components merely by signs, and so the components of the tensor all transform via the one dimensional representations of $\grp{Q}_8$, which are equivalent to representations of $\grp{K}_4$. From a representation theoretic point of view one has the decomposition:
\begin{equation}
    \mathbf{S} \otimes \mathbf{S} = 4(\mathbf{1}_1)+4(\mathbf{1}_P)+4(\mathbf{1}_T)+4(\mathbf{1}_{PT}).
\end{equation}
One can read off assignments of different Klein group representations to the various gamma matrices in Tab.~(\ref{tab:q8decomp}). 
\begin{table}[h]
    \centering
    \begin{tabular}{|cc|c|}\hline
        Basis & Component &  $\grp{K}_4$ Rep \\\hline
         $\mathbb{I}_4, \>\>\gamma^i\gamma^j$ & $p_0, \>\> v_i$ & $1$ \\
         $\gamma^0, \>\>\gamma^i \gamma^5$ & $s, \>\> G_{ij}$ &  $T$ \\
          $\gamma^i, \>\>\gamma^0 \gamma^5$ & $G_{0i}, \>\> a$ &  $P$ \\
         $\gamma^5, \>\>\gamma^0 \gamma^i$ & $v_0, \>\> p_i$ & $PT$ \\\hline
    \end{tabular}
    \caption{The decomposition of the $\mathbf{S} \otimes \mathbf{S}$ representation of the pinoral discrete reflection group $\grp{Q}_8$. Notice one really only needs the assignments of the $\gamma^\mu$ to complete the rest.}
    \label{tab:q8decomp}
\end{table}

\noindent
We may put this information together more compactly: let us refer to the component symbols as the vectors and tensors themselves, with the understanding that the sum is within the Clifford algebra, and use a subscript to indicate the charge of the objects under the discrete reflection group $\grp{K}_4$:
\begin{equation}
    \psi \phi^\intercal = s_T + p_{T} + G_{T} + v_P + a_{P}
\end{equation}
In order we have a $T$-odd pseudo scalar, a momentum type four-vector, an anti-symmetric tensor behaving like an angular momentum tensor, a polarization type vector, and a $P$-odd pseudo scalar. 

Now we can also see why we were careful about the introduction of the adjoint spinor $\overline{\phi}:= \phi^\intercal \gamma^0$. The `spinor metric' $\gamma^0$ is charged under the reflection group as $T$, so if we are not careful when using it, it is easy to make a mistake. Let us see how the adjoint affects the various fermion bi-linears. Taking the tensor product instead with the adjoint gives:
\begin{equation}
\label{Eq:standardDiracDecomp}
    \psi \otimes \overline{\phi} = \psi \phi^\intercal \gamma^0 = s \mathbb{I}_4 + p_\mu \gamma^\mu + G_{\mu \nu} \gamma^\mu \gamma^\nu + v_\mu \gamma^\mu\gamma^5 + a \gamma^5,
\end{equation}
the standard, more satisfying decomposition. How does the presence of $\gamma^0$ affect a time reversal transform? Let's inspect
\begin{equation}
(\mathbf{T}\psi) \otimes \overline{(\mathbf{T}\phi)}= \mathbf{T}\left(\psi \otimes \phi^\intercal\right) \mathbf{T}^\intercal \gamma^0 = -\mathbf{T}\left(\psi \otimes \phi^\intercal \gamma^0\right) \mathbf{T}^\intercal.
\end{equation}
Before even finishing the computation, in the middle we can see the time reversal would be carried out exactly as in Eq.~(\ref{eq:Q8PtensorDecomp}), and then simply rearranged by $\gamma^0$ back to the standard decomposition in Eq.~(\ref{Eq:standardDiracDecomp}). We can check explicitly by still carrying out the rest of the computation:
\begin{equation}
\begin{split}
    -\mathbf{T}\left(\psi \otimes \phi^\intercal \gamma^0\right) \mathbf{T}^\intercal &= -\mathbf{T} \left(s \mathbb{I}_4 + p_\mu \gamma^\mu + G_{\mu \nu} \gamma^\mu \gamma^\nu + v_\mu \gamma^\mu\gamma^5 + a \gamma^5\right)\mathbf{T}^\intercal \\
    &=-s \mathbb{I}_4 + p_0 \gamma^0 - p_i\gamma^i + G_{0i} \gamma^0 \gamma^i - G_{ij} \gamma^i \gamma^j - v_0 \gamma^0\gamma^5 + v_i \gamma^i\gamma^5 + a \gamma^5.
\end{split}
\end{equation}
The time reversal operator acts as one would expect on the Clifford algebra bases, followed by an additional sign. This is exactly what happens in the classical case \cite{ConvLorRep} when combining tensors of different type: just as the angular momentum tensor holds onto the extra $T$ charge from the momentum vector, the tensor decomposition of $\psi \otimes \overline{\phi}$ remembers the $T$ charge of $\gamma^0$. Thus while one might anticipate from commutation relations alone that the $\gamma^\mu$ components of the tensor decomposition should be a coordinate ($1$-type) vector, in fact they form a momentum ($T$-type) vector. 

The conclusion of the above is that it is useful to consider there being four fundamental representations of $\Pin{3,1}$, labeled $1, P, T, PT$ in accord with the vector case. The only difference being that the $\mathbf{P}$, $\mathbf{T}$, and $\mathbf{PT}$ operators act irreducibly on the spinor components. Importantly, the discrete reflection operators in the $\mathbf{S}$ representation of $\grp{Q}_8$ are themselves charged under the reflection group! Of course they must be in order for the group to be non-Abelian, and Tab.~(\ref{tab:q8decomp}) tells us precisely how they are charged: $\mathbf{P}$ is charged under $T$, $\mathbf{T}$ is charged under $P$, and $\mathbf{PT}$ is charged under $PT$. One simple takeaway from this is that a generic fermion current
\begin{equation}
    \phi^\intercal \gamma^0 \widehat{O} \psi,
\end{equation}
will be charged under the discrete reflection group as $T \cdot K[O]$ where $K$ returns the charge of $\widehat{O}$.

\newpage

\subsection{\texorpdfstring{$\grp{D}_8$ Representation Theory for $\Pin{1,3}$}{D8 Representation Theory for pin(1,3)}}
\begin{table}[!h]
    \hspace{-2em}\Centering
    \begin{tabular}{|c|cccc|}\hline
        Name & $\pm 1$ & $\pm \mathbf{P}$ & $\pm \mathbf{T}$ & $\pm \mathbf{PT}$ \\\hline
        $\rho_1$ & $1$ & $1$ & $1$ & $1$ \\
        $\rho_T$ & $1$ & $1$ & $-1$ & $-1$ \\
        $\rho_P$ & $1$ & $-1$ & $1$ & $-1$ \\
        $\rho_{PT}$ & $1$ & $-1$ & $-1$ & $1$ \\ \hline
        $\rho_S$ & $\pm\begin{pmatrix}
            1 & 0 \\
            0 & 1  \\
        \end{pmatrix}$ & $\pm\begin{pmatrix}
            1 & 0 \\
            0 & -1  \\
        \end{pmatrix}$ & $\pm\begin{pmatrix}
            0 & 1 \\
            1 & 0  \\
        \end{pmatrix}$& $\pm\begin{pmatrix}
            0 & -1 \\
            1 & 0  \\
        \end{pmatrix}$ \\\hline
    \end{tabular}
    \caption{Irreducible Representations of $\grp{D}_8$ over $\C$, or equivalently $\R$. The one dimensional representations are labeled by which elements act as $-1$, and $\rho_S$ the `spinor' rep, is the largest and only possible real faithful irrep.}
    \label{tab:D8Reps}
\end{table}
\noindent
The story for $\Pin{1,3}$ is essentially the same as $\Pin{3,1}$. The tensor decomposition plays out exactly analogous to Eq.~(\ref{eq:genericdecomp}) and Eq.~(\ref{Eq:standardDiracDecomp}), where the quaternion conjugate transpose is taken everywhere instead of transpose; the analogously named Clifford algebra elements in $\Pin{1,3}$ have the same reflection charges as in Tab.~(\ref{tab:q8decomp}). This makes sense as when go from the pin groups to vector representations of $\Orth{1,3}$, both pin groups ought to deliver us the same representations. 

\section{\texorpdfstring{$\mathcal{CPT}$}{CPT} and Co-representations: Acquiring Anti-linearity}
\label{sec:CPTcorep}
One way to understand why charge conjugation ($\mathcal{C}$), should be intertwined with parity ($\mathcal{P}$) and time reversal ($\mathcal{T}$), is by understanding it as arising from a particular central extension of the Lorentzian pin groups: in particular this extension is the symmetry of $\mathcal{CPT}$. Generally, any Lie group with real or quaternionic fundamental representations will necessarily have a symmetry operation given by an anti-linear intertwiner, and so then have a `natural' central extension by $\Z_2$, essentially including the relevant anti-linear intertwiner as an element of a centrally extended group. Representations which map group elements not only to linear maps but also anti-linear maps are known as co-representations (see Appendix.~(\ref{app:introcorep}) for a review of co-representations, and Appendix.~(\ref{sec:CoRepZ2Ext}) for building them as central extensions of the aforementioned kind). In even dimensions all pin groups may be extended this way to introduce an operator which commutes with all spacetime symmetries. In odd dimension, one of the two pin groups may be constructed explicitly as a central extension of this kind from the spin group. In either case, one finds a natural way of not only finding the $\mathcal{CPT}$ operator for any spacetime, but also of realizing time reversal as an anti-linear operator. 

\noindent These constructions by definition would be trivial were one to remain in the real or quaternionic settings; a real or quaternionic structure can only act non-trivially in the complexified setting. One may ask then why do this at all? If our pin groups are naturally real or quaternionic, why enter the complex setting and extend the group by the `spurious' discrete symmetry, instead of staying in the setting where it is manifest? The answer is that particle states and fields in spacetime take representations not only in the relevant spacetime symmetry groups, but also representations of the relevant gauge groups, which generally are understood to be complex (the $\SU{3}$ symmetry of chromodynamics for example has fundamentally complex representations). As such, linearity of the tensor product between spacetime and gauge representations brings us to the complex setting, both by observation and convenience. In what follows an anti-linear operator will be denoted by the corresponding matrix part, and a $*$, indicating complex conjugation waiting to act to the right.

\subsection{\texorpdfstring{${\mathcal{CPT-}}$extension of $\grp{Pin}(3,1)$}{CPT-extension of Pin(3,1)}}
From the above exposition, the $\mathcal{CPT-}$extension of our real (Majorana) pin group is done by simply extending the symmetries operators at hand by complex conjugation $*$, which obviously commutes with every action of the given representation of $\Pin{3,1}$. Since conjugation squares to the identity, the extended group is equivalent to $\Pin{3,1} \times \Z_2$. Let us refer to the extended group as $\Pin[\star]{3,1}$. The purpose of this extension of course is to build a natural group structure where we find a time reversal operator acting anti-linearly, and squaring to $-1$. The sixteen elements of our extended discrete reflection group $\grp{Q}_8 \times \Z_2$ are
\begin{equation}
\begin{split}
    &\gamma^0, \quad \gamma^5 \gamma^0, \quad \gamma^0 *, \quad \gamma^5 \gamma^0 *,\\
    &\mathbb{I}_4,\> \qquad\gamma^5, \quad\> \mathbb{I}_4*, \qquad \gamma^5 *,
\end{split}
\end{equation}
as well as their negations. The top line is all those elements which anti-commute with generators of boosts, and so span our options for $\mathcal{P}$ and $\mathcal{T}$. The bottom line are all elements which commute with the action of the spin subgroup, and so span our options for $\mathcal{PT}$. If $\mathcal{P}$ is to be linear, and $\mathcal{T}$ anti-linear, then $\mathcal{PT}$ must be anti-linear. Furthermore to keep the structure of the $\Spin{3,1}$ subgroup Eq.~(\ref{eq:fullSpin}) in tact, $\mathcal{PT}$ must square to $-1$. This restricts us to precisely $\mathcal{PT} = \pm\gamma^5 *$. This couples our choices of $\mathcal{P}$ and $\mathcal{T}$ as in the linear case: choosing $\mathcal{P} = \pm\gamma^0$, forces us to have $\mathcal{T} = \pm\gamma^5\gamma^0*$, while choosing $\mathcal{P} = \pm\gamma^5\gamma^0$, demands $\mathcal{T} = \pm\gamma^0*$. Taking the same choice as above $\mathcal{P} = \mathbf{P}$, we have the following updated assignment of operators:
\begin{equation}
    \mathcal{P} = \gamma^0, \quad \mathcal{T} = \gamma^5\gamma^0 *, \quad \mathcal{PT} = \gamma^5 *, \quad \mathcal{CPT} = \mathbb{I}_4*.
\end{equation}
We have an additional generator of the extended group $\Pin[\star]{3,1}$, innocently called $\mathcal{CPT}$. We will discuss and defend this name in a later section. It is simple to confirm in this case we still have that $\mathcal{P}$, $\mathcal{T}$, and $\mathcal{PT}$ form $\grp{Q}_8$, and so the representation theory of the tensor product decompositions is unchanged. Importantly $\mathcal{T}^2 = -1$, and so the time reversal operator has no eigenvalues or eigenvectors even over $\C$, as one expects for time reversal on spinors. Note that now, should imaginary numbers enter we have that they are charged under time reversal: $K[i] = T$. With this extended group structure, let us now change the basis of our $\spin{3,1}$ generators to that of the standard Weyl (chiral) basis. With a change of basis matrix
\begin{equation}
    U^{-1} = \frac{1}{\sqrt{2}} \begin{pmatrix}
        1 & i & 0 & 0 \\
        0 & 0 & 1 & i \\
        0 & 0 & -i & -1 \\
        i & 1 & 0 & 0 
    \end{pmatrix},
\end{equation}
our generators, reflections, and spinor metric $h = \gamma^0$ will transform via:
\begin{equation}
\gamma^\mu \mapsto U^{-1} \gamma^\mu U, \quad \mathcal{T} \mapsto U^{-1} \mathcal{T} U^**, \quad h \mapsto U^\dagger h U.
\end{equation}
$\mathcal{P}$ transforms like the $\gamma^\mu$ matrices, while $\mathcal{PT}$ and $\mathcal{CPT}$ transform like $\mathcal{T}$. With all symbols explicitly in the Weyl basis our operators may be written:
\begin{equation}
    \mathcal{P} = \gamma^0 ,\quad \mathcal{T} =  \gamma^1 \gamma^3 * , \quad 
    \mathcal{PT} = -\gamma^5 \gamma^2 * , \quad \mathcal{CPT} = - \gamma^2 *, \quad h = \gamma^0.
\end{equation}

\subsection{\texorpdfstring{${\mathcal{CPT-}}$extension of $\grp{Pin}(1,3)$}{CPT-extension of pin(1,3)}}
In the quaternionic setting, the $\mathcal{CPT}-$extension is given by including the operator $j\>\widetilde{\left(\>\right)}^*$, the quaternionic intertwiner,\footnote{See Sec.~(A.1.2) of \cite{QuatOrth} for a review of this quaternion-reversion-conjugate operation.} in this notation waiting to act to the right. This squares to $-1$ and so the extended group has the structure $\Pin[\star]{1,3} = \left(\Pin{1,3} \times \Z_4\right)/\Z_2$. We may wish to learn from the above and simply tack this intertwiner onto the linear version of our operators, since by definition it commutes with all operations and so will leave the commutation behaviour in tact. Unfortunately we need to be more careful than this. The elements of our extended discrete reflection group\footnote{This group is isomorphic to the so called `Pauli Group': the group generated by the three standard Pauli matrices.} $\left(\grp{D}_8 \times \Z_4 \right)/\Z_2$ are 
\begin{equation}
\begin{split}
    &\gamma^0, \quad \gamma^5 \gamma^0, \quad \gamma^0 j\>\widetilde{\left(\>\right)}^*, \quad \gamma^5 \gamma^0 j\>\widetilde{\left(\>\right)}^*,\\
    &\mathbb{I}_4,\> \qquad\gamma^5, \quad\> \mathbb{I}_4 j\>\widetilde{\left(\>\right)}^*, \qquad \gamma^5 j\>\widetilde{\left(\>\right)}^*,
\end{split}
\end{equation}
and their negations. Again the top row are all those elements which invert boosts, and the bottom row are those which commute with the spin group. Following identical logic to the previous case, if $\mathcal{P}$ is linear and $\mathcal{T}$ anti-linear, than $\mathcal{PT}$ must be anti-linear. For the structure of $\Spin{1,3} \equiv \Spin{3,1}$ to be the same, $\mathcal{PT}$ must square to $-1$. This is where our analysis diverges.

$\mathcal{PT}$ must be central, anti-linear, and square to $-1$, and this forces to have $\mathcal{PT} = \pm j\>\widetilde{\left(\>\right)}^*$. From this we have that either $\mathcal{P} = \pm\gamma^0$ and $\mathcal{T} = \pm \gamma^0 j\>\widetilde{\left(\>\right)}^*$, or $\mathcal{P} = \pm\gamma^5\gamma^0$ and $\mathcal{T} = \pm \gamma^5\gamma^0 j\>\widetilde{\left(\>\right)}^*$. Taking the former option gives:
\begin{equation}
\begin{split}
    \mathcal{P} = \gamma^0, \quad \mathcal{T} = \gamma^0 j \>\widetilde{()}^* , \quad \mathcal{PT} = j \>\widetilde{()}^*, \quad \mathcal{CPT} &= \gamma^5 j \>\widetilde{()}^* .
\end{split}
\end{equation}
In the interest of preserving the structure of the spin group within our extended group, we are forced to have $\mathcal{P}$ and $\mathcal{T}$ operators which commute with one another, altering our discrete reflection group. Where previously we had the structure of the group $\grp{D}_8$, we now find the structure of a new group: $\mathcal{P}$, $\mathcal{T}$ and $\mathcal{PT}$ here square to $+1$, $-1$, $-1$ respectively. It is not hard to show this is equivalent to the group $\Z_4 \times \Z_2$. This group is abelian, and another double covering of $\grp{K}_4$. 

Since we find ourselves with a modified discrete reflection group we will need to revisit our tensor product decomposition. The abelian nature of the group simplifies our life a great deal, as $\mathcal{P}$ and $\mathcal{T}$ are no longer charged under the reflection group. In this case we can see for an arbitrary decomposition:
\begin{equation}
\begin{split}
    \mathcal{T}\left[\psi \phi^\dagger \gamma^0 \right] &= \left(\mathcal{T}\psi\right) \left(\mathcal{T}\phi\right)^\dagger \gamma^0  = \gamma^0 j\>\widetilde{\psi}^* \widetilde{\phi}^\intercal (-j) \gamma^0 \gamma^0 = \gamma^0 j \widetilde{\left(\psi \widetilde{\phi}^\dagger \gamma^0\right)^*} (-j) \gamma^0 \\
    &= \gamma^0 \left(\psi \widetilde{\phi}^\dagger \gamma^0\right)\gamma^0 = \mathcal{P}\left[\psi \phi^\dagger \gamma^0 \right]\\
    & =\gamma^0 \left( s \mathbb{I}_4 + p_0 \gamma^0 + p_i\gamma^i + G_{0i} \gamma^0 \gamma^i + G_{ij} \gamma^i \gamma^j + v_0 \gamma^0\gamma^5 + v_i \gamma^i\gamma^5 + a \gamma^5 \right) \gamma^0\\
    &= s \mathbb{I}_4 + p_0 \gamma^0 - p_i\gamma^i - G_{0i} \gamma^0 \gamma^i + G_{ij} \gamma^i \gamma^j - v_0 \gamma^0\gamma^5 + v_i \gamma^i\gamma^5 - a \gamma^5 
\end{split}
\end{equation}
Written more compactly the above tells us:
\begin{equation}
    \psi \overline{\phi}  = s_1 + p_{T} + G_{1} + v_P + a_{PT}.
\end{equation}
Interestingly, this tensor product takes us from $\Pin[\star]{1,3} \mapsto \PO{1,3}$, i.e. $\mathcal{PT}$ acts as the identity on the tensor product space. This makes sense as $\mathcal{PT}$ is defined as the intertwiner commuting with the entire $\cl{1,3}$ algebra, in stark contrast to the $\Pin[\star]{3,1}$ case. It bears repeating the above in words: just because the reflection group is abelian, does not mean the elements of the Clifford algebra are not charged under the reflection group. Importantly $\gamma^5$ is still charged under $PT$. 

\noindent In order to bring this quaternionic setting to the complex one, we need to map both the quaternion vectors to complex vectors, and the quaternion matrices to complex matrices. One way this may be done is as follows. Replace every quaternion vector with one of twice the dimension, stacking element wise. Then as well, replace every quaternion matrix entry blockwise:
\begin{equation}
\label{eq:qVtoCV}
    \begin{pmatrix}
        t+ix+jy+kz  \\
        \vdots
    \end{pmatrix} \mapsto \begin{pmatrix}
        t + i z \\
        y + i x \\
        \vdots
    \end{pmatrix}, \qquad 
    t+ix+jy+kz   \mapsto \begin{pmatrix}
        t + iz & ix-y \\
        ix + y & t - iz
    \end{pmatrix}.  
\end{equation}
Once this is done we may go to the `same'\footnote{Formally, these are different groups and so the basis cannot meaningfully be `the same'. However we can map so their $\spin{1,3}$ algebra's are represented identically, and then impose that the expressions of $\gamma^\mu$ in either case differ by multiplication of $i$. This choice is not unique however, one could also choose $\gamma^\mu_- = i \gamma^5 \gamma^\mu_+$} Weyl basis as we did above, via a change of basis 
\begin{equation}
    U^{-1} = \frac{1}{\sqrt{2}}\begin{pmatrix}
        0 & -i & 0 & 1 \\
        i & 0 & -1 & 0 \\
        0 & -i & 0 & -1 \\
        i & 0 & 1 & 0 
    \end{pmatrix}.
\end{equation}
Giving the operators expressed in the new basis as
\begin{equation}
\begin{split}
    \mathcal{P} = \gamma^0 ,\quad \mathcal{T} = \gamma^1 \gamma^3 * , \quad 
    \mathcal{PT} = \gamma^5\gamma^2 *, \quad \mathcal{CPT} = -\gamma^2 *, \quad h = \gamma^0.\\
\end{split}
\end{equation}
Which look a great deal familiar. Let us contrast these results with the $\Pin[\star]{3,1}$.

\subsection{Comparison of the two \texorpdfstring{$\mathcal{CPT}-$}{CPT-}extended pin groups}
Starting with both extended pin groups in their respective Weyl bases, we will refer to elements of $\Pin[\star]{3,1}$ with a $+$ subscript, and $\Pin[\star]{1,3}$ with a $-$. In terms of differences between the groups, we can see the following holds between the bases:
\begin{equation}
\begin{split}
    \gamma^\mu_- &= i \gamma^\mu_+, \qquad\>\>\>\> \gamma^5_- = \gamma^5_+, \qquad\qquad \>\>\mathcal{P}_- = i \mathcal{P}_+, \\
    \mathcal{T}_- &= - \>\mathcal{T}_+, \quad \mathcal{PT}_- = i\>\mathcal{PT}_+, \quad \mathcal{CPT}_- = i\> \mathcal{CPT}_+.
\end{split}
\end{equation}
Despite the disanalogies in the setup of the respective cases, the operators all only differ by a phase. Firstly we should note the analogous roles of $\mathcal{PT}$ and $\mathcal{CPT}$: in either case both commute with the action of $\Spin[+]{1,3}$, however $\mathcal{CPT}_+$ commutes with the entire $\Pin[\star]{3,1}$ group, while $\mathcal{PT}_-$ commutes with the entire $\Pin[\star]{1,3}$.\footnote{One could cutely refer to these as the spintertwiner's and the pintertwiner's.} 

\noindent Another relationship is easy to see: in either case we have from the anti-linearity of time reversal $K[i] = T$. This fact helps connect the two pin groups in the complex setting. The difference between the `spinor metric' in the two cases is a multiplication by $i$, and the same for the $\gamma^\mu$. Importantly in $\Pin[\star]{3,1}$ we have
\begin{equation}
\begin{split}
    \psi \phi^\dagger \gamma^0 &= s_T + p_{T} + F_{T} + v_P + a_{P}\\
    &= s\mathbb{I}_4 + p_\mu \gamma^\mu + F_{\mu\nu} \gamma^\mu \gamma^\nu + v_\mu \gamma^5 \gamma^\mu + a \gamma^5.
\end{split}
\end{equation}
Swapping signatures here is then essentially the same as putting an $i$ onto every $\gamma^\mu$, including the spinor metric. This has the overall effect of adding a $T$ charge on the even elements of the decomposition: the scalar, pseudo-scalar, and tensor pieces. This changes the charges of the decomposition to become: $s_1 + p_{T} + F_{1} + v_P + a_{PT}$, which is precisely the $\Pin[\star]{1,3}$ decomposition. The differences between the decomposition of the pin groups raises the question of whether or not parity violation could be understood as an interaction between two different kinds of pinors. This will be considered in future work.

\subsubsection{Tensor Products}
One issue with the promotion of our real and quaternionic pinors to the complex setting, is that the tensor product does not `commute' with the complexification. Take for example an arbitrary $-$ pinor $\varphi$ in the explicitly quaternionic setting of $\Pin{1,3}$. In this setting the outer product of the spinor with itself has the following decomposition:
\begin{equation}
    \varphi \overline{\varphi} = s \mathbb{I}_2 + p_\mu \gamma^\mu + a \gamma^5
\end{equation}
Where $s, p_\mu, a$ are all real, and the tensor and axial-vector components are zero. However if we map the pinor into the complex setting $ \Hq^2 \ni \varphi \mapsto \psi \in \C^4$ via the prescription of Eq.~(\ref{eq:qVtoCV}) the decomposition picks up additional terms, all of which have imaginary coefficients:
\begin{equation}
    \psi \overline{\psi} = \left(s \mathbb{I}_4 + p_\mu \gamma^\mu + a \gamma^5\right) + i (F_{\mu\nu} \gamma^\mu \gamma^\nu + v_\mu \gamma^5 \gamma^\mu)
\end{equation}
We can isolate these real and imaginary parts by utilizing the so called pintertwiner, in this case $\mathcal{PT}$, as the spurious imaginary units will be odd under $\mathcal{PT}$:
\begin{equation}
\begin{split}
    \mathcal{PT}\left[\psi \overline{\psi}\right] &= \left(\mathcal{PT} \psi^*\right) \left(\mathcal{PT} \psi^*\right)^\dagger \gamma^0 = \mathcal{PT} \left(\psi ^* \psi^\intercal \gamma^0\right) \mathcal{PT}^\dagger = \mathcal{PT} \left(\psi \overline{\psi}\right)^* \mathcal{PT}^\dagger \\
    &= \mathcal{PT}\left(\left(s \mathbb{I}_4 + p_\mu \gamma^\mu + a \gamma^5\right) + i \left(F_{\mu\nu} \gamma^\mu \gamma^\nu + v_\mu \gamma^5 \gamma^\mu\right)\right)^* \mathcal{PT}^\dagger\\
    &=\left(s \mathbb{I}_4 + p_\mu \gamma^\mu + a \gamma^5\right) - i \left(F_{\mu\nu} \gamma^\mu \gamma^\nu + v_\mu \gamma^5 \gamma^\mu\right).
\end{split}
\end{equation}
Thus the relationship between tensor products in the quaternionic setting to the complex setting is
\begin{equation}
     \varphi \overline{\varphi} \mapsto \frac{1}{2}\left(\psi \overline{\psi} + \psi_{\mathcal{PT}} \overline{\psi}_{\mathcal{PT}}\right),
\end{equation}
allowing us to respect the extended $\Pin[\star]{1,3}$ symmetry. However the imaginary piece here also respects the $\Pin[\star]{1,3}$ symmetry if we simply divide out the $i$. That is to say, in the complex setting we have interesting new bi-linears
\begin{equation}
    \frac{1}{2i}\left(\psi \overline{\psi} - \psi_{\mathcal{PT}} \overline{\psi}_{\mathcal{PT}}\right) = F_{\mu\nu}\gamma^\mu\gamma^\nu + v_{\mu}\gamma^5\gamma^\mu
\end{equation} 
and so in the complex setting of $\Pin[\star]{1,3}$ there are additional physical fermion bi-linear's for axial currents and anti-symmetric tensors given by $\overline{\psi} \gamma^\mu i\gamma^5\psi$ and $\overline{\psi} \gamma^\mu \gamma^\nu i\psi$. The presence of the $i$ alters the time reversal properties of the bi-linear, each picking up a $T$ charge. More generally a tensor product in the complex setting may be split into these real and imaginary parts, giving a whole other set of possible bi-linear's: 
\begin{equation}
    \psi_1 \overline{\psi_2} = \left(s_1+s_T\right) + \left(p_T+p_1\right) + \left(F_1+F_T\right) + \left(v_P+v_{PT}\right) + \left(a_{PT}+a_P\right)
\end{equation}
where each pair is a real-imaginary pair. All this together tells us the physical bi-linear's we can extract from a single complex $\Pin[\star]{1,3}$ pinor are:
\begin{equation}
    \psi \overline{\psi} = s_1+p_T+F_T+v_{PT} + a_{PT}
\end{equation}

\noindent The story is similar for the `Majorana' pin group $\Pin[\star]{3,1}$: a manifestly real pinor $\phi$ has the following decomposition:
\begin{equation}
    \phi \overline{\phi} = p_\mu \gamma^\mu + G_{\mu\nu} \gamma^\mu \gamma^\nu.
\end{equation}
When we perform a complex change of basis, the pinor may appear generically complex. However it can be seen to still be a $\Pin[\star]{3,1}$ pinor because it will remain an eigenvector of the transformed $\mathcal{CPT}_+$ operator with eigenvalue $+1$. In this case the decomposition will not be altered. However, if we intentionally complexify by say, demanding a $\U{1}$ symmetry, a generic complex pinor transforming under $\Pin[\star]{3,1}$ will have a decomposition of
\begin{equation}
    \phi \overline{\phi} = \left(p_\mu \gamma^\mu + G_{\mu\nu} \gamma^\mu \gamma^\nu \right) +i\left(s \mathbb{I}_4 + v_\mu \gamma^5 \gamma^\mu + a \gamma^5\right),
\end{equation}
where each component can be extracted with proper insertion of $i$'s into the bi-linears, or via splitting the bilinear into $\mathcal{CPT}_+$ odd and even parts, analogously to above. A generic tensor product decomposes into 
\begin{equation}
    \psi_1 \overline{\psi_2} = \left(s_T+s_1\right) + \left(p_T+p_1\right) + \left(F_T+F_1\right) + \left(v_P+v_{PT}\right) + \left(a_{P}+a_{PT}\right)
\end{equation}
where again each pair is a real-imaginary pair. All this together tells us the physical bi-linear's we can extract from a single complex $\Pin[\star]{3,1}$ pinor are:
\begin{equation}
    \psi \overline{\psi} = s_1+p_T+F_T+v_{PT} + a_{PT}
\end{equation}
Which are precisely the same observables from $\Pin[\star]{1,3}$! On their own, once the pinor's are brought to the complex setting the difference between them is moot. However this does not preclude the possibility of interactions between the two kinds of pinors. It is note worthy that both cases have an  imaginary axial vector, motivating the introduction of an $i$ to the definition of $\gamma^5$ for convenience in the complex setting, though the psudeo-scalar is only imaginary in the Majorana case. In future work the author intends to study the behaviour of tensors built from one pinor of each type, to see how the decomposition changes.

\subsection{A Defense of the Name of the \texorpdfstring{$\mathcal{CPT}$}{CPT} Operator}
There is a great deal of inconsistency in language between many sources on the topic of anti-matter and charge conjugation. The above sections imply an odd identification
\begin{equation}
    \mathcal{C} = \mathcal{CPT} \>\left(\mathcal{PT}\right)^{-1} = \gamma^5,
\end{equation}
which on the surface appears wrong. Yet if $\mathcal{P}$ is to be linear, with $\mathcal{T}$ and $\mathcal{CPT}$ anti-linear, there is not a reasonable definition of $\mathcal{C}$ which is anti-linear, despite what is written in most textbooks. What most textbooks call $\mathcal{C}$ or $\mathcal{CP}$ must in fact contain a time reversal if the operations are indeed anti-linear, and it is this desire to have $\mathcal{C}$ be `whatever operator relates particles and anti-particles' which is the source of the confusion felt by many students and even experts in the subject. This assignment of the $\mathcal{C}$ operator is more defensible than it seems: the standard spinor solutions of the Dirac equation, typically denoted by: $u(p)^\uparrow$, $u(p)^\downarrow$, $v(p)^\uparrow$, $v(p)^\downarrow$, may all be related via:
\begin{equation}
    u(p)^\downarrow= \mathcal{PT} \> u^*(p)^\uparrow, \quad v(p)^\uparrow = \mathcal{CPT} \>u^*(p)^\uparrow, \quad v(p)^\downarrow = \mathcal{C}\> u(p)^\uparrow,
\end{equation}
notice the anti-particle polarization states are related by $\mathcal{C}$ and $\mathcal{CPT}$.\footnote{The author presently is still wrestling with how precisely to think of anti-matter, and its relationship to advanced solutions and the Feynman-Stückelberg interpretation. Presently it seems as though defining anti-matter as the $\mathcal{CPT}$ conjugate of an advanced state is the only generic possibility in semi-classical field theory.}

Charge conjugation is confusing because it is something to be done at the level of assignments of representations: it is the action by the physicist of declaring that for every fields representation in the theory, they will now instead be assigning the conjugate representation. This operation is linear because it does not descend to the underlying field anti-linearly. The effect of charge conjugation on say, a quark field transforming under the $\mathbf{3}$ representation of $\SU{3}$, is to declare that all gauge transformations are performed instead as if the quark were charged under $\overline{\mathbf{3}}$. In particular it does so linearly, not affecting phases $\mathcal{C}[e^{i \phi}\psi_\mathbf{3}] \mapsto e^{i \phi} C\psi_\mathbf{\bar{3}}$. Here $C$ must be some matrix that commutes with $\Spin[+]{3,1}$. Notice no conjugation occurs. It might perhaps be more comfortable to reconcile this fact by recalling that for all unitary Lie groups, the conjugate representation of the Lie algebra is isomorphic to the dual representation, and the dual representation is found by taking the negative transpose of every element, which is a linear operation. Charge conjugation might be more aptly referred to as charge duality, to avoid the implication of complex conjugation. It is still the anti-unitarity of time reversal which gives rise to its entangling with $\mathcal{C}$, as $\mathcal{T}$'s anti-linearity alters the gauge transform behaviour of advanced solutions of wave equations, but the $\mathcal{C}$ itself must be linear to make sense. This assignment is also interesting seeing as $\grp{SU}(2)$ is the only gauge symmetry group in the standard model whose representations are self-conjugate, possibly leading to an interpretation of the left hand projection operators serving not to project out a certain chirality, but to project to eigen-states under $\mathcal{C}$. This will be explored in future work.
 
\subsection{A Curious Equivalence of Majorana and Weyl spinors}
\label{sec:majasweyl}
Above a co-representation of $\Pin{3,1}$ was built by a $\mathcal{CPT}-$extension, and we called the extended group $\Pin[\star]{3,1}$. In this particular case we have another option: squashing our Clifford algebra from four real dimensions to two complex dimensions. To fit our above description of $\Pin{3,1}$ into operators on $\C^2$ instead of on $\R^4$, we have to make a decision regarding that map; in particular which reflection within the Clifford algebra is mapped to complex conjugation. From Eq.~(\ref{eq:CL31basis}) an obvious choice is $\gamma^2$, which implies the map from real spinors to complex spinors
\begin{equation}
    \begin{pmatrix}
        a \\
        b\\
        c\\
        d
    \end{pmatrix} \rightarrow \begin{pmatrix}
        a + i b \\
        c + i d
    \end{pmatrix}.
\end{equation}
This alternating choice is especially convenient as it is much easier to identify `real' and `complex' blocks of real matrices via the identification \begin{equation}
    1 \leftrightarrow \bb{I}_2, \quad i \leftrightarrow \begin{pmatrix}
        0 & -1 \\
        1 & 0 
    \end{pmatrix}.
\end{equation}
Let us see this explicitly for our generators of $\spin{3,1}$:
\begin{equation}
L_x = \frac{1}{2}\begin{pmatrix}
 0 & 0 & 0 & -1 \\
 0 & 0 & 1 & 0 \\
 0 & -1 & 0 & 0 \\
 1 & 0 & 0 & 0 \\
\end{pmatrix}, \quad L_y = \frac{1}{2}\begin{pmatrix}
 0 & 0 & -1 & 0 \\
 0 & 0 & 0 & -1 \\
 1 & 0 & 0 & 0 \\
 0 & 1 & 0 & 0 \\
\end{pmatrix}, \quad L_z = \frac{1}{2}\begin{pmatrix}
 0 & -1 & 0 & 0 \\
 1 & 0 & 0 & 0 \\
 0 & 0 & 0 & 1 \\
 0 & 0 & -1 & 0 \\
\end{pmatrix},
\end{equation}

\noindent 
It is simple to see these are `block-complex', and can be mapped to
\begin{equation}
S_x = \frac{1}{2}\begin{pmatrix}
 0 & i \\
 i & 0 \\
\end{pmatrix}, \quad S_y = \frac{1}{2}\begin{pmatrix}
 0 & -1 \\
 1 & 0 
\end{pmatrix}, \quad S_z = \frac{1}{2}\begin{pmatrix}
i & 0 \\
 0 & -i \\
\end{pmatrix}.
\end{equation}
The boosts are similarly mapped as 
\begin{equation}
B_x = \frac{1}{2}\begin{pmatrix}
 0 & 1 \\
 1 & 0 \\
\end{pmatrix}, \quad B_y = \frac{1}{2}\begin{pmatrix}
 0 & i \\
 -i & 0 
\end{pmatrix}, \quad B_z = \frac{1}{2}\begin{pmatrix}
1 & 0 \\
 0 & -1 \\
\end{pmatrix}.
\end{equation}
Importantly the center of $\Spin{3,1}$ is mapped to
\begin{equation}
    \gamma^5 \mapsto \begin{pmatrix}
        i & 0 \\
        0 & i 
    \end{pmatrix}.
\end{equation}
I.e. $\mathbf{PT}$ is the scalar $i$. This gives a fun identification of 
\begin{equation}
    \Spin{3,1} = \SL[\pm]{2, \C}_{|\R}.
\end{equation}
The full spin group (both components) is given by those complex matrices with determinant $\pm 1$. As for the reflections $\mathbf{P}$ and $\mathbf{T}$, the parity operator is mapped as follows:
\begin{equation}
    \gamma^0 = \gamma^0 \gamma^2 \gamma^2 = \begin{pmatrix}
 0 & 0 & 0 & 1 \\
 0 & 0 & -1 & 0 \\
 0 & -1 & 0 & 0 \\
 1 & 0 & 0 & 0 \\
\end{pmatrix}\gamma^2 \quad \mapsto\quad \begin{pmatrix}
    0 & -i \\
    i &0 
\end{pmatrix}*,
\end{equation}
where the complex conjugation is waiting to act to the right. It should be the case that, at least up to a sign, $\mathbf{T} = \mathbf{P} \cdot \mathbf{PT}$. We can check this going to the other way (remembering the conjugation is an operator, and so should sticks around waiting to act):
\begin{equation}
    \mathbf{T} = \begin{pmatrix}
    0 & -i \\
    i &0 
\end{pmatrix}\left(i\right)^* *= \begin{pmatrix}
    0 & -1 \\
    1 & 0 
\end{pmatrix} * := \varepsilon *
\end{equation}
Where $\varepsilon$ is a name for the given matrix. When mapped back into the $4\times 4$ case this can be seen to be proportional to $\gamma^5\gamma^0$. All in all we have the assignment:
\begin{equation}
    \mathbf{P} = \begin{pmatrix}
        0 & -i \\
        i & 0 
    \end{pmatrix}*\>, \quad \mathbf{T} = \begin{pmatrix}
        0 & -1 \\
        1 & 0 
    \end{pmatrix}*\>, \quad \mathbf{PT} = i.
\end{equation}
Notice the complex-conjugation is what allows the operators of $\mathbf{P}$, $\mathbf{T}$, and $\mathbf{PT}$ to mutually anti-commute. We have successfully squashed down our Majorana spinors to Weyl spinors, but at the cost of having a $\mathbf{P}$ which is anti-linear, and a $\mathbf{PT}$ which is linear. An alternative to this is explored in Appendix.~(\ref{app:worse}), though it is unlikely to be fruitful. Importantly this is a homomorphism to $\Pin{3,1}$ and so the decomposition of a tensor product of a Majorana pinor into a four vector anti-symmetric tensor should still hold. For a Weyl spinor $\psi$:
\begin{equation}
    \psi\psi^\dagger = E \mathbb{I}_2 + p_i \sigma^i  \quad \Rightarrow \quad \psi^\dagger\sigma^\mu \psi = p^\mu
\end{equation}
It can be seen the $p_i$ components are odd under $\mathbf{P}$ and $\mathbf{T}$, giving a $T-$type four vector, the momenta. The anti-symmetric tensor is missing. It has been sent to the linear tensor product:
\begin{equation}
    \psi\psi^T \varepsilon = F_{0i}\> \sigma^i + F_{ij} \>\sigma^i \sigma^k.
\end{equation}
Here the $F_{0i}$ are $P$-odd and $F_{ij}$ are $T$-odd. In fact as $\mathbf{PT}$ is the action of a phase, to find any elements which are $PT$-odd the product cannot be unitary. In this two dimensional complex case, these are the only possible bi-linear constructions, and so the mapping is curious, because it truly is in exactly $1$ to $1$ correspondence with the Majorana case. The only disanalogy is in the standard treatment of Weyl spinors as components of Dirac bi-spinors. Promoting to the Majorana case gives distinctly non-chiral spinors, whereas mapping to the Dirac case gives explicitly chiral spinors. One wonders if there is an equivalent formulation of the weak interaction using these Majorana-Weyl equivalent pinors. This will be explored in future work.

\subsection{\texorpdfstring{Co-representation of $\grp{Pin}(1,4)$}{Co-representation of pin(1,4)}}
The de Sitter group $\Orth{1,4}$ is the symmetry group of de Sitter spacetime (four dimensional spacetime with constant positive spacial curvature), or equivalently as the non-translational symmetries of a $1+4$ Minkowski spacetime. The group is ten dimensional, and thought of in the context of de Sitter spacetime, the extra group generators compared to the Lorentz group are not extra directions to rotate through or boost into, but serve as compact and non-compact translations through the positively curved spacetime. The de Sitter spacetime is characterized by a de Sitter radius which all translation parameters are relative to.

From the classification of Clifford algebra's one might assume we should to start building matrices in $\M{4}{\C}$ or $\grp{M}^2_2\left({\Hq}\right)$, however in this case we will start with the same Clifford algebra and bases as $\Pin{1,3}$, because we know that $\M{4}{\C} \cong \left(\M{2}{\Hq} \times \Z_4\right) /\Z_2 $, meaning our $\mathcal{CPT}$-extension of $\Pin{1,3}$ should also suffice for building a co-representation of $\Pin{1,4}$. One main difference in the consideration of these groups however, is that in this case $\gamma^5$, which squares to $-1$, is explicitly included as a fundamental generator of the Clifford algebra. Doing so extends our spin Lie algebra from the Lorentzian case: not only are the usual second order elements included, but what in the Lorentzian case were tri-vectors, are now also second order products. The group generators are 
\begin{equation}
    s^{AB} = \frac{1}{2}\gamma^A \gamma^B ,\quad A, B  \in\{ 0,1,2,3,5 \}.
\end{equation}
The first $6$ generators are the standard boosts and rotations. $\gamma^0\gamma^5$ squares to $1$ and so should generate a non-compact translation, namely time translation, while $\gamma^i\gamma^5$ are compact and generate something like a rotation, in this case interpreted as rotation around the compact spacial dimensions of the positively curved de Sitter universe, which locally appears as translation. It is often useful to discuss the group equivalently thought of as $5$ dimensional Minkowski spacetime, in which case one simply imagines we have named the coordinate axes $0,1,2,3,5$, skipping over four. All in all from the original basis for $\Pin{1,3}$, with $L$ for rotations, $K$ for boosts, and $T$ for translations, the generators of $\spin{1,4}$ in this quaternionic setting can be seen to be:
\begin{equation}
\begin{split}
    \begin{Bmatrix}
        L_x \\
        L_y \\ 
        L_z 
    \end{Bmatrix} &= \frac{1}{2}     \begin{Bmatrix}
        i \\
        j \\ 
        k 
    \end{Bmatrix} \begin{pmatrix}
        1 & 0 \\ 
        0 & 1
    \end{pmatrix} ,\quad \begin{Bmatrix}
        K_x \\
        K_y \\ 
        K_z 
    \end{Bmatrix} = \frac{1}{2}     \begin{Bmatrix}
        i \\
        j \\ 
        k 
    \end{Bmatrix} \begin{pmatrix}
        0 & 1 \\ 
        -1 & 0
    \end{pmatrix}, \\
    \begin{Bmatrix}
        T_x \\
        T_y \\ 
        T_z 
    \end{Bmatrix} &= \frac{1}{2}\begin{Bmatrix}
        i \\
        j \\ 
        k
    \end{Bmatrix} \begin{pmatrix}
        1 & 0 \\
        0 & -1
    \end{pmatrix}, \qquad 
    \>\>T_t = \frac{1}{2}\begin{pmatrix}
        0 & 1 \\
        1 & 0
    \end{pmatrix}.
\end{split}
\end{equation}
Firstly it should be noted that $\gamma^0$ is still a Hermitian structure upon these elements: $\lambda ^\dagger\gamma^0 = -\gamma^0\lambda $ where $\lambda$ are the Lie algebra elements. This means $\gamma^0$ may still serve as the adjoint matrix $h$. Playing the same game as for the Lorentzian pin group, we search for some elements of the Clifford algebra which behave as we desire for $\mathbf{P}$ and $\mathbf{T}$ operators, namely: $\mathbf{P}$ should commute with rotations, reverse boosts, and reverse spacial translations, commuting with time translation. On the other hand, $\mathbf{T}$ should commute with rotations, reverse boosts, and reverse only time translation, commuting with spacial translation. The constraints from rotations and boosts bring us to the same choices as for the above Lorentzian pin group, and of those options it can be seen only $\gamma^0\gamma^5$ could serve as $\mathbf{P}$, and $\mathbf{T}$ must be $\gamma^0$.

One might be surprised we started with a group that was too small and yet, before even including anti-linearity, found seemingly all the right ingredients for the Clifford algebra of de Sitter space. The trick to this is that what we have built above is not a faithful double covering of $\Orth{1,4}$, but of $\PO{1,4} \cong \SO{1,4}$. That is we have a linear representation of the group $\Spin{1,4}$ only. Importantly notice the composition of the following `rotations'
\begin{equation}
    \exp{\left( \pi L_{z} \right)} = \gamma^1 \gamma^2, \quad \exp{\left( \pi \>T_{z} \right)}= \gamma^3 \gamma^5,
\end{equation}
gives $-\gamma^0$, and so $\gamma^0$ is not a reflection in this construction, even though if it appears to be behaving as one. $\gamma^0$ should be thought of as the action of spacial inversion, distinct from parity in even spacial dimensions. To understand this better, note the structure of the de Sitter group $\Orth{1,4}$ is not quite the same as the Lorentz group $\Orth{1,3}$. While they both have four connected components and a center of $\Z_2 = \pm 1$, importantly the Lorentz group's non trivial central element is $\mathcal{PT}$, while for the de Sitter group the non trivial central element is contained in the $T$ component of the group, in particular as shown above it is two commuting `rotations' by $\pi$ away from $\mathcal{T}$:
\begin{equation}
\begin{split}
    \mathcal{T} &= \grp{diag}\{-1,1,1,1,1\} \\ 
    &= \grp{diag}\{1,-1,-1,1,1\} \cdot \grp{diag}\{1,1,1,-1,-1\}\cdot  \grp{diag}\{-1,-1,-1,-1,-1\} \\
    &= R_{12}( \pi ) R_{35}\left(\pi\right) \left(-\mathbb{I}_5\right).
\end{split}
\end{equation}
What this means is that were $\mathbf{T} = \gamma^0$ mapped into the vector case, it is acting not as $\grp{diag}\{-1,1,1,1,1\}$, but by the element it is related to by application of the center: $\grp{diag}\{1,-1,-1,-1,-1\}$, spacial inversion. The present construction then is a double covering of the centerless orthogonal group $\PO{1,4}$, where all spacial reflections behave as expected, but all time reversals behave with an additional sign, shrinking the group down to only two of its four disconnected components. 

We still expect the correct time reversal operator $\mathcal{T}$ in the full group to be anti-linear, and so to acquire the entire Pin group, we will extend this formulation by the quaternionic intertwiner $j \widetilde{\>\>\>}^*$, which we will see this time is a construction worthy of the name $\Pin{1,4}$. The elements from which we may build our discrete reflection group are below
\begin{equation}
\begin{split}
    &\gamma^0, \quad \gamma^5 \gamma^0, \quad \gamma^0 j\>\widetilde{\left(\>\right)}^*, \quad \gamma^5 \gamma^0 j\>\widetilde{\left(\>\right)}^*,\\
    &\mathbb{I}_2,\> \qquad\gamma^5, \quad\> \mathbb{I}_2 j\>\widetilde{\left(\>\right)}^*, \qquad \gamma^5 j\>\widetilde{\left(\>\right)}^*.
\end{split}
\end{equation}
This is thus far identical to the case of $\Pin[\star]{1,3}$, except for the definiteness in the assignment of $\mathbf{P}$. Of these only $\gamma^0 j\>\widetilde{\left(\>\right)}^*$ has the correct behaviour to be $\mathcal{T}$: it reverses the correct generators, is anti-linear, and squares to $-1$.

As a sanity check of the covering homomorphism, we can ask what element double covers the non-trivial central element of $\Orth{1,4}$? Well, it should be a spacial inversion away from $\mathcal{T}$, which is simply $j\>\widetilde{\left(\>\right)}^*$, which is indeed central, and in the same component as $\mathcal{T}$. This also all means that $\mathcal{PT} = \gamma^5 j\>\widetilde{\left(\>\right)}^*$, which was a problem in the Lorentzian case because it had the wrong square. However it is only a rotation away from $\gamma^0 \gamma^5 j\>\widetilde{\left(\>\right)}^*$, and so concerns about its square not being $-1$ are a matter of convention at worst. Conversely, basic orthogonality arguments show $\gamma^5$ is still not path connected to the identity, and so can still serve as a reflection, bringing us between group components. All in all the following assignment of operators allow for the construction of a co-representation of $\Pin{1,4}$
\begin{equation}
    \mathcal{P} = \gamma^0 \gamma^5, \quad \mathcal{T} = \gamma^0 j\>\widetilde{\left(\>\right)}^*, \quad \mathcal{PT} = \gamma^5 j\>\widetilde{\left(\>\right)}^*, \quad \mathcal{CPT} = j\>\widetilde{\left(\>\right)}^*.
\end{equation}
This assignment gives the standard behaviour which extends the Lorentzian assignments of parity and time reversal charges to the de Sitter case. In this case the discrete reflection group remains $\grp{D}_8$. Let us see how this affects our pinor behaviour. Under parity and time reversal we have:
\begin{equation}
\begin{split}
    \mathcal{P}\left[\psi \overline{\varphi}\right] &= \mathcal{P}\psi \varphi^\dagger\mathcal{P}^\dagger \gamma^0\\
    &=-\gamma^0 \gamma^5 \left(s \mathbb{I}_4 + p_0 \gamma^0 + p_i\gamma^i + G_{0i} \gamma^0 \gamma^i + G_{ij} \gamma^i \gamma^j + v_0 \gamma^0\gamma^5 + v_i \gamma^i\gamma^5 + a \gamma^5 \right)\gamma^0 \gamma^5 \\
    &= \left(-s \mathbb{I}_4 + p_0 \gamma^0 - p_i\gamma^i + G_{0i} \gamma^0 \gamma^i - G_{ij} \gamma^i \gamma^j - v_0 \gamma^0\gamma^5 + v_i \gamma^i\gamma^5 + a \gamma^5 \right)\\
    \mathcal{T}\left[\psi \overline{\varphi}\right] &= \mathcal{T}\psi \varphi^\dagger\mathcal{T}^\dagger \gamma^0\\
    &=\gamma^0 \left(s \mathbb{I}_4 + p_0 \gamma^0 + p_i\gamma^i + G_{0i} \gamma^0 \gamma^i + G_{ij} \gamma^i \gamma^j + v_0 \gamma^0\gamma^5 + v_i \gamma^i\gamma^5 + a \gamma^5 \right)\gamma^0 \\
    &= \left(s \mathbb{I}_4 + p_0 \gamma^0 - p_i\gamma^i - G_{0i} \gamma^0 \gamma^i + G_{ij} \gamma^i \gamma^j - v_0 \gamma^0\gamma^5 + v_i \gamma^i\gamma^5 - a \gamma^5 \right)
\end{split}
\end{equation}
If we wish to be consistent, the representations here need to be understood as objects transforming under the de Sitter group $\Orth{1,4}$, of which the relevant representations are the $1$ dimensional representation, the $5$ dimensional vector representation, and the $10$ dimensional adjoint (anti-symmetric tensor) representation. Labeling by dimension and reflection charges, the de Sitter decomposition would be:
\begin{equation}
    \psi \overline{\varphi} = \mathbf{1}_P+\mathbf{5}_T+\mathbf{10}_{P}.
\end{equation}
In the large de Sitter radius limit, the de Sitter representations decompose as $\mathbf{5} \mapsto \mathbf{4}\oplus \mathbf{1}$ and $\mathbf{10} \mapsto \mathbf{6} \oplus \mathbf{4}$, which results a Lorentzian decomposition of:
\begin{equation}
    \psi \overline{\varphi} = s_P+p_T+G_P+v_P+a_T.
\end{equation}
This situation is potentially alarming if the full de Sitter symmetry is to be respected, as it means there are no true scalar de Sitter mass terms, only psudeo-scalar. Even considering the promotion to the equivalent complex formulation, extra imaginary terms would give another scalar which is $PT$-odd, not a true scalar either. This invites the possibility of a theory which is formally a massless theory in de Sitter spacetime, which locally looks like a massive Lorentzian theory.\footnote{This is even more interesting when considering that conformal theories are one of the few types which evade the Coleman-Mandula theorem, which normally forbids the mixing of gauge symmetries and spacetime symmetries.} This will be studied in future work. 

\section{Summary}
From the broadest perspective, this work justifies the standard treatment of the parity and time reversal operators in standard Dirac theory. More specifically spinoral representations of the two Lorentzian pin groups were constructed, analyzed, and contrasted in detail. These pin groups were then extended to co-representations in order to accommodate the anti-linearity demanded by any reasonable time reversal operator. Doing so in either case, we found an additional discrete symmetry identified with the operation of $\mathcal{CPT}$, relating particles and anti-particles. 

Perhaps the most important result shown was how once either kind of pinor enters the complex domain, the differences in the choice of pin group become moot, as all resulting bi-linear currents have the same properties under $\Orth{1,3}$. It was also shown how Weyl spinors may be embedded in two completely different ways to the two kinds of pinors: they may be embedded in the standard way as handed Dirac spinors, or they may be promoted to Majorana spinors. A quaternionic representation of the de Sitter pin group $\Pin{1,4}$ was built, and it was shown in this framework there are no truly scalar mass terms, only psudeo-scalar.

In future work the author intends to probe the possible relationship between the various pin groups and the weak nuclear force. In particular I intend to determine whether the handed nature of the weak force may be constructed in any of the following ways: as a combination of two different kinds of Pinors, or recast in terms of the equivalent Majorana form of the Weyl spinors, or understood formally as combinations of currents which respect foremost a $\mathcal{C}$-symmetry, with the parity violation stemming essentially from the self-conjugacy of the symmetry group of flavourdynamics. The author also intends to take seriously the consideration that a massless fermionic theory in de Sitter spacetime, in the limit of large de Sitter radius may appear locally as a Lorentzian theory of massive fermions. 

\newpage 
\appendix

\section{Introduction to co-representations}
\label{app:introcorep}
Co-representations\footnote{Before learning of their construction and use by Wigner, I ended up inventing co-representations myself and named them $*-$representations. For this reason the appendix may be overly pedagogical as it was written before I had found other references on the topic.} fall out a necessity to handle the physically motivated observation that certain transformations upon vector spaces (changes of frame) should be defined not as linear maps, but as anti-linear maps: the composition of a linear map with complex conjugation. Particularly regarding time reversal within the Lorentz group, standard representation theory would promise to map all our group elements to linear transformations upon any given state space. This means the standard approach to the representation theory of the Lorentz group cannot fully capture the mathematics required to talk about space-time reflections. The demand that time reversal be represented by an anti-linear transform begs for the construction of an apparatus to do so: a co-representation is just that. A co-representation of a group is, for the most part, a representation in the usual sense, however the action of the group upon the relevant state-space may be realized as either a linear operator or anti-linear operator. Further motivation and a formal definition are given below.
\subsection{Motivation}
Firstly let us see why standard `linear' representation theory may leave one wanting, time reversal aside for now. One motivation for the introduction of co-representations is when working with real-representations of a group which has a center containing the Klein group $\mathrm{K}_4$. In this case any and all irreducible representation (irrep) maps will have non-trivial kernel. That is to say there are no faithful irreps: to study representations one either looks for faithful reducible representations, or unfaithful irreps. We are familiar with unfaithful irreps, most adjoint actions and covering maps are representations of this kind: however it is uncomfortable to find one is in a situation where there are \textit{no} irreps which serve to faithfully represent the entire group; this is famously the case for $\Spin{8}$. Below we will explore how a co-representation may remedy this discomfort, `disguising' a reducible real representation of even dimension into something resembling an irrep instead over the complex numbers in half the dimension. Of course will come to find the concept useful for more than merely groups of this type. 

The primitive example of a co-representation is that of the Klein group $\grp{K}_4 = \Z_2 \times \Z_2$. Being an Abelian group, by Schur's lemma we know that every \textit{irreducible} representation over $\C$ must be one dimensional, and every irrep must map the center of the group to scalar multiples of the identity. Given the defining relations of the Klein group:
\begin{equation}
    \grp{K}_4 = \{1, a,b,c \quad | \quad a\cdot b = c, \quad a^2 = b^2 = c^2 = 1\},
\end{equation}
we have that all of our elements square to the identity, and all our elements commute, and so the center of the group is the entire group. Thus in order to have a faithful irrep we would need four distinct second roots of unity over $\C$ --- but of course there are only $\pm1$. So the smallest \textit{faithful} linear representation then is two dimensional. As the roots of two are both real, we may represent the Klein group acting upon $\R^2$ via the following representation $\rho$:
\begin{equation}
    \rho(1) = \begin{pmatrix}
        1 & 0  \\
        0 & 1
    \end{pmatrix}, \quad \rho(a) = \begin{pmatrix}
        -1 & 0  \\
        0 & -1
    \end{pmatrix}, \quad \rho(b) = \begin{pmatrix}
        0 & 1  \\
        1 & 0
    \end{pmatrix}, \quad \rho(c) = \begin{pmatrix}
        0 & -1  \\
        -1 & 0
    \end{pmatrix}.
\end{equation}
Clearly this representation is faithful, as each element is mapped to a unique matrix; but this means it must therefore be reducible. This may not be obvious at first glace, but as all four of these matrices commute, they can be made to act irreducibly (simultaneously block diagonally) on a shared eigenbasis, which decomposes the space into irreps. This change of basis gives
\begin{equation}
    \rho^\prime(1) = \begin{pmatrix}
        1 & 0  \\
        0 & 1
    \end{pmatrix},\quad \rho^\prime(a) = \begin{pmatrix}
        -1 & 0  \\
        0 & -1
    \end{pmatrix}, \quad \rho^\prime(b) = \begin{pmatrix}
        1 & 0  \\
        0 & -1
    \end{pmatrix}, \quad \rho^\prime(c) = \begin{pmatrix}
        -1 & 0  \\
        0 & 1
    \end{pmatrix},
\end{equation}
decomposing the action of $\mathrm{K}_4$ on $\R^2$ into an action on $\R \oplus \R$. 

Desirous of something resembling a faithful irrep, and we may try to eat our cake and have it to by identifying $\R^2$ with $ \C$ via the map $\sigma$:
\begin{equation}
\sigma:\R^2 \rightarrow \C \quad \mathrm{via}\quad \begin{pmatrix}
    x \\
    y
\end{pmatrix}    \mapsto x + i y.
\end{equation}
Under this identification, how can we map the action of the Klein group \textit{faithfully}? The first operations are easy, as the identity matrix and its negation can be mapped from $\grp{GL}(\R^2)$ into $\grp{GL}(\C^1) \cong \C^\times$ as $\pm 1$. However $\rho^\prime (b)$ and its negation have no analogue in $\grp{GL}(\C^1)$. Applying $\rho^\prime (b)$ on $\R^2$ and then applying $\sigma$, it is clear this operation is the action of complex conjugation! Instead of giving up because this is not a linear representation (and not complex differentiable), we will simply map it as such as press on. We may say $\mathrm{K}_4$ has a faithful $1$ dimensional co–representation $\rho^\star$, given by:
\begin{equation}
    \rho^\star(1)(x+iy) = 1, \quad \rho^\star(a)(x + iy) = -1, \quad \rho^\star(b) = \>^*,\quad \rho^\star(c) = -\>^*,
\end{equation}
or for clarity:
\begin{equation}
\begin{split}
    \rho^\star(1)(x+iy) = x + i y,&\quad \rho^\star(a)(x + iy) = -x-i y,\\
    \quad \rho^\star(b)(x+iy) = x-i y,&\quad \rho^\star(c)(x+iy) = -x+ iy. 
\end{split}
\end{equation}
Another way to say this is that the co-representation is the composition of maps:
\begin{equation}
    \rho^\star = \sigma \rho^\prime \sigma^{-1}.
\end{equation}
It is easy to verify these operations still form the Klein group under composition, and so we still have a group homomorphism to some set of operations on $\mathbb{C}$. Let us define this below.

\subsection{Definition}
Co-representations are formally a particular kind of \textit{semi-linear} representation. For any field (or division ring) $\bb{K}$ with outer automorphism group $\grp{Out}(\bb{K})$, the \textit{general semi-linear group} denoted $\grp{\Gamma L}$ on an $n$ dimensional $\bb{K}$-vector space $V = \mathbb{K}[e_1, \ldots, e_n]$ is defined as the following semi-direct product:
\begin{equation}
    \grp{\Gamma L}(V) = \grp{GL}(V) \rtimes \grp{Out}(\bb{K}).
\end{equation}
Where in the automorphisms act naturally on components. If all automorphisms are inner, this is the same as the general linear group. For $\bb{K} = \C$, the outer automorphism group is populated only by complex conjugation: $\grp{Out(\C)} =\Z_2 = \{1, \>^*\>\}$; as we are focused on this case will name this particular general semi-linear group, the `star general linear group':
\begin{equation}
    \grp{GL}^{\star}(\C^n) := \grp{\Gamma L}(\C^n) \sim \GL{\C^n} \oplus \overline{\grp{GL}}(\C^n).
\end{equation}
This is the group of all linear transforms on $\C^n$ --- of course known as $\GL{\C^n}$ --- as well as all \textit{anti-linear} transforms on $\C^n$ --- named $\overline{\grp{GL}}(\C^n)$ --- of which all elements may be written as a complex conjugation followed by an element of $\GL{\C^n}$. Note the anti-linear piece, $\overline{\grp{GL}}(\C^n)$ is a component of the group not connected to the identity component of $\grp{GL}^\star(\C^n)$, and so is not itself a subgroup. We are now ready define what we mean in the first section by co-representation.

\subsubsection*{Formal Definition}
Given a group $G$, a co-representation of $G$ on $\C^n$ is a group homomorphism from $G$ into $\grp{GL}^\star(\C^n)$. That is to say, a map $\rho$ is a co-representation when
\begin{equation}
    \{\rho: G \mapsto \grp{GL}^\star(\C^n)\quad | \quad \rho(gh) = \rho(g)\rho(h) ,\quad \forall g,h, \in G\}.
\end{equation}
I.e. the image of $\rho$ is either a linear map $A \in \grp{GL}(\C^n)$ or may be written as the composition of a linear map and complex conjugation: $A \circ *$. 

As outer automorphisms are generally not differentiable maps, one may wonder what role they can play in regards to Lie group theory. The appeal of co-representations is that they can give convenient realizations of maps between \textit{disconnected} components of a Lie group, describing for example, reflections, which are also not differentiable maps within the Lie group.

\subsection{Dimension Counting}
As vector spaces alone there is no question that one can always map complex vectors to real ones of twice the dimension and vice versa: $\C^n \leftrightarrow \R^{2n}$. In particular one can (non-canonically) map pairs of real basis ($e_i$) to a single complex one ($f_i$) as was done above with the map $\sigma$: 
\begin{equation}
    e_1 \mapsto f_1  \quad  e_2 \mapsto i f_1.
\end{equation}
However if one is interested in generic linear transformations of these vectors, then the representation theory of the general linear group on these spaces are not analogous. In general we have that
\begin{equation}
     \grp{dim}_\R\grp{GL}(\C^n) = 2 \grp{dim}_\C\grp{GL}(\C^n) = 2 n^2, \qquad \grp{dim}_\R\grp{GL}(\R^{2n}) = (2n)^2 = 4 n^2.
\end{equation}
So $\GL{\C^n}$ can be neatly embedded inside $\grp{GL}(\R^{2n})$, but not the other way around. In particular from a complex matrix we may embed non-canonically into a real matrix of twice the dimension by the standard identifications:
\begin{equation}
\label{eq:CtoR}
    1 \leftrightarrow \bb{I}_2, \quad i \leftrightarrow \begin{pmatrix}
        0 & -1 \\
        1 & 0 
    \end{pmatrix}.
\end{equation}
However going the other way is not always possible on a strictly linear basis, as we saw above with the Klein group. The matrix 
\begin{equation}
     \begin{pmatrix}
        1 & 0  \\
        0 & -1
    \end{pmatrix}
\end{equation}
cannot be mapped to any strictly linear map on the complex vector space $\C$. As we have seen above the key to understanding this disanalogy is complex conjugation. Many of the maps in $\GL{\R^{2n}} $, when composed with $\sigma$, will correspond to a linear transform \textit{and} a complex conjugation of the corresponding vector in $\C^n$, and more generally the action of these matrices could be decomposed into a sum of a linear and anti-linear maps. 
\subsection{\texorpdfstring{$\Z_2$ Grading of $\GL{\R^{2n}}$}{Z2 Grading of GL(2n, R)}}
What the above is getting at is that every even dimensional general linear group over the reals, when thought of equivalently as a subset of its matrix algebra, may be considered to be $\Z_2$ graded: its elements split into two parts. It is no problem to see the grading of $\grp{GL}^\star(\C^{n})$: the composition of two $\C$-linear transforms will again be $\C$-linear. The composition of two anti-linear transformations will involve two complex conjugations, giving us something $\C$-linear. And the composition of an anti-linear transform with a linear transform will be anti-linear. Thus the elements of $\GL{\R^{2n}}$ can be assigned a parity corresponding to whether they can be mapped to a linear map, or anti-linear map. 

The grading at the level of $\grp{GL}(\R^{2n})$ is less obvious, but can be constructed conveniently as follows. Consider $\grp{GL}(\R^{2n})$ as a subalgebra of the matrix algebra $\grp{M}_{2n}(\R)$, with the group and algebra products identified. This is just the invertible elements of $\grp{M}_{2n}(\R)$. Conveniently we can identify this real even dimensional matrix algebra as the Clifford algebra $\cl{n+1,n-1}$, and so we can identify $\grp{GL}(\R^{2n})$ as the group of invertible elements of this Clifford algebra. Clifford algebras are naturally $\Z_2$ graded, so this point of view provides us precisely the structure we are after. There are other Clifford algebras (different possible signatures) we could choose which would span the same matrix algebra; this choice of signature was made because for all $n$, the even subalgebra of $\cl{n+1,n-1} \cong \M{2n}{\R}$, is precisely 
\begin{equation}
\label{eq:even2complex}
    \grp{C\ell}^0(n+1,n-1) \cong \cl{n+1,n-2} \cong \M{n}{\C},
\end{equation}
and the invertible elements of $\M{n}{\C}$ may be identified with the elements of $\GL{\C^{n}}$. Thus elements of $\GL{\R^{2n}}$ via the natural embedding into the invertible elements of $\cl{n+1,n-1}$, can be decomposed into even and odd elements of the Clifford algebra. The even subspace can be mapped (non-canonically) to a linear action on a complex vector space of half the dimension via Eq.~(\ref{eq:even2complex}). Elements of the odd subspace must be mapped as anti-linear transforms. In particular if we look at the odd elements of the Clifford algebra, one can make a choice of one element $K =\operatorname{diag}\{+1,-1,\ldots,+1,-1\}$ to be the group element identified with complex conjugation.\footnote{Of course any similarity transform away from this works as well. The matrix merely needs to be traceless with eigenvalues of $\pm$1 (the matrix squares to the identity, just as conjugation does).} This choice is defined by, and equivalently defines, the particular choice of map from $\R^{2n} \mapsto \C^n$. Once this choice is made, every element of the odd subspace can be factored as an element of the even subalgebra times $K$. Since we demand the map from $\GL{\R^{2n}} \mapsto \GL[\star]{\C^n}$ be a homomorphism, the even piece of these odd elements are mapped as above, and $K$ is mapped to complex conjugation. Thus every element of good parity in $\GL{\R^{2n}}$ can be mapped to an element of $\GL[\star]{\C^n}$, and we have at least one example of how one can meaningfully and consistently define a `group representation' which maps some elements of the group to anti-linear operators.

We will see this in action in the next section. It should be noted that this decomposition and mapping will be extremely convenient for groups which naturally come out of this Clifford algebra approach, like spin groups, whereas the same approach for say orthogonal groups, which can sit inside even dimensional the general linear groups just as well but do not have such a natural graded structure, require a more precise the notion of `adding' linear and anti-linear maps to realize the group homomorphism. 

\section{\texorpdfstring{$\mathcal{CPT}$}{CPT}: Co-representations of Central Extensions by \texorpdfstring{$\Z_2$}{Z2}}
\label{sec:CoRepZ2Ext}
Within the complex setting, real and quaternionic representations may be identified as follows. For a real Lie group $G$ acting on $\C^n$ via some irreducible representation map $\rho: G \mapsto \GL{n,\mathbb{C}}$, suppose we have an anti-linear map $\sigma$ acting upon the same representation space. A representation is self-conjugate if there exists a $\sigma$ whose action commutes with every group action upon the space. More precisely, for any scalars $a,b \in \C$, any vectors $v, w \in \C^n$, and the representation of any group element $R \in \rho(G)$ we have that $\sigma$, when it exists, can always be chosen so that:
\begin{equation}
    \sigma(Rv) = R \sigma(v), \quad \mathrm{and} \quad \sigma(av+bw) = \overline{a} \sigma(v)+ \overline{b}\sigma(w), \quad \mathrm{and} \quad \sigma(\sigma(v)) = \pm v.
\end{equation}
When such an intertwiner exists it falls into one of two classes, it is either a real structure($+$) of a quaternionic structure($-$).
\subsection{\texorpdfstring{$\Z_2$}{Z2} central extension of a Real Representation}
When $\sigma^2 = 1$, the representation is a real representation, and one can always find a manifestly real basis wherein $\sigma$ acts merely as complex conjugation,\footnote{Under a change of basis matrix $U$, if the linear piece of $\sigma$ is given by $\Sigma$, then in the new basis $\sigma$ will be given by $\sigma^\prime = U^{-1}\Sigma U^* *$.} and the representation maps all matrices into $\GL{\R^n}\subset\GL{\C^n}$. When one has a real representation, a central extension of $G$ by $\Z_2$: $G\times \Z_2$, may have a realization upon $\C^n$ as a co-representation $\rho^\star$ via the obvious extension:
\begin{equation}
    \rho^\star: G\times \Z_2 \cong (G, 0) \oplus(G, 1)\mapsto \left(\rho\right) \>\oplus\> \left(\rho \circ \sigma\right).
\end{equation}
Essentially the new central element is mapped to the commuting real structure, for which there is some basis where it is merely complex conjugation. 

\subsection{\texorpdfstring{$\Z_2$}{Z2} central extension of a Quaternionic Representation}
When $\sigma^2 = -1$, the representation is a quaternionic representation. In this case the representation must be even dimensional $n = 2m$. We will rename $\sigma$ to $J$ here, reserving the former for real representations only. In this case, one can always find a basis where $i$, $J$, and $iJ$ can be mapped to the quaternions $i,j,k$ respectively, and the representation map is restricted to $\GL{\Hq^m} \subset \GL{\C^n}$. In this quaternionic setting $J$ may be realized as the trivial intertwiner $j \>\widetilde{\left(\>\>\right)}^*$ (where $*$ is quaternion conjugation and $\widetilde{\>\>\>\>}$ is quaternion reversion. For a detailed summary of these and review of quaternion groups and self-conjugacy see \cite{QuatOrth} by the author). When one has a quaternionic representation, extending the group to include the action of $J$ is not quite as simple as the real case, because $J^2 = -1$ must be mapped to the same $-1$ contained within $G$. This is accomplished via the following non-trivial central extension of $G$ by $\mathbb{Z}_2$:
\begin{equation}
    \left(G\times \Z_4\right)/\Z_2.
\end{equation}
This should be read as extending the group by something generating a $\Z_4$ (in this case $J$), and then identifying the $-1$ within each group to be the same element. In some sense this may be regarded as a kind of `discrete complexification' of $G$, in the sense that $\Z_4$ can be modeled as fourth roots of unity and so generated by $i$. This central extension has a co-representation $\rho^\star$ via:
\begin{equation}
    \rho^\star: G\times \Z_2 \cong (G, \pm 1) \oplus(G, \pm i)\mapsto \left(\rho_r\right) \>\oplus\> \left(\rho_r \circ J\right).
\end{equation}
Essentially the new central element is mapped to the intertwining quaternionic structure. 

\subsection{Odd Dimensional Pin Groups as \texorpdfstring{$\mathcal{CPT}-$}{CPT-}extensions of Spin groups}
While the Lorentz group is even dimensional, understanding the relationship between it, and related constructions such as the Euclidean pin groups in three dimensional, and the de Sitter pin groups in $4+1$ dimensions, can help enlighten us regarding the nature of $\mathcal{C}$, $\mathcal{P}$, and $\mathcal{T}$. Important relations between odd dimensional pin and spin groups are collected in Tab.~(\ref{tab:PinExtendSpin}). 
    
Notice swapping the signature swaps the roles of $p$ and $q$, interchanging the given central extension, so the two pin groups are seen to be the two different central extensions of $\grp{Spin}(p,q)$ by $\Z_2$. This relationship is precisely the same structure seen above of taking a real or quaternionic representation, and extending the group to include the commuting anti-linear structure. Thus in cases where odd dimensional $\Spin{p,q}$ representations are real or quaternionic, one of the pin groups will naturally as this anti-linear extension of the spin group. As it turns out, every finite dimensional representation of the spin group in odd dimensions will be either real or quaternionic, and so from the point of view of co-representations, one pin group is always `more natural' to construct. 

\begin{table}[!h]
    \hspace{-2em}\Centering
    \begin{tabular}{|ccc|}\hline
        Signature & $\quad$ As a $\Z_2$ extension of $\Spin{p,q}$ & Center of $\Pin{p,q}$ \\\hline
        $p-q = 1 \>\operatorname{mod} \> 4$ & $\Pin{p,q} \cong \Spin{p,q} \times \Z_2$ & $\Z_2 \times \Z_2$ \\
        $p-q = 3 \>\operatorname{mod}\> 4$ & $\qquad\>\>\>\Pin{p,q} \cong \left(\Spin{p,q} \times \Z_4\right)/ \Z_2$ & $\Z_4$ \\\hline
    \end{tabular}
    \caption{For $p$ $+$ $q$ $=$ $n$ odd, the pin groups may be realized as central extensions of the relevant spin group by $\Z_2$. As every spin group in odd dimension is real or quaternionic, one of the two pin groups may be considered as the natural central extension by the corresponding real or quaternionic anti-linear structure in the defining representation.}
    \label{tab:PinExtendSpin}
\end{table}

\section{Experimenting with Co-representations}
\subsection{\texorpdfstring{Example: Co-representation of $\Pin{3,0}$ as an extension of $\Spin{3}$}{Example: Co-representation of pin(3,0) as an extension of spin(3)}}
\subsubsection*{Linear Pin Group}
\label{sec:su2}
Famously we have the `accidental isomorphisms': $\Spin{3}\cong \SU{2}\cong \Sp{1}$. In the standard description of $\Pin{3,0}$ inside the Clifford algebra $\cl{3,0} \cong \M{2}{\C}$, the generators of the algebra are taken to be the Pauli matrices:
\begin{equation}
    \sigma_x = \begin{pmatrix}
        0 & 1 \\
        1 & 0
    \end{pmatrix}, \quad \sigma_y = \begin{pmatrix}
        0 & -i \\
        i & 0
    \end{pmatrix}, \quad \sigma_z = \begin{pmatrix}
        1 & 0 \\
        0 & -1
    \end{pmatrix}.
\end{equation}
and the subalgebra $\spin{3}$ is spanned by order two elements of this Clifford algebra $\sigma_i \sigma_j$ for which one basis is:
\begin{equation}
\label{eq:su2basis}
    s_{xy} = \frac{1}{2}\begin{pmatrix}
        0 & -i \\
        -i & 0
    \end{pmatrix}, \quad s_{zx} = \frac{1}{2}\begin{pmatrix}
        0 & -1 \\
        1 & 0
    \end{pmatrix}, \quad s_{xy} = \frac{1}{2}\begin{pmatrix}
        -i & 0 \\
        0 & i
    \end{pmatrix}.
\end{equation}
The remaining units of the Clifford algebra are given by $\mathbb{I}_2$ and $i\mathbb{I}_2$, forming the center of $\Pin{3,0}$, and completing the basis of this $2^3 = 8$ real-dimensional algebra. 

From Tab.~(\ref{tab:PinExtendSpin}), we can see $\Pin{3,0} \cong \left(\Spin{3}\times \Z_4\right)/\Z_2$. Interpreting this using standard linear representation theory is tantamount to the observation that $\Pin{3,0}$ is given by $\Spin{3}$ along with $i$ times $\Spin{3}$. Geometrically $\Pin{3,0}$ thought of in terms of performing rotations and reflections, can be understood as having rotations given as usual by $\Spin{3}$, and all reflections may be acquired via some rotation multiplied by $i$, understood as parity, total spacial inversion. This gives a convenient matrix identification of the group: 
\begin{equation}
    \Pin{3,0} = \{S \in \grp{U}(2) \>| \> \det S = \pm 1  \} := \SU[\pm]{2}.
\end{equation}

\subsubsection*{Anti-linear Pin Group}
The Clifford algebra approach tells us that the volume element $\omega$ of $\Pin{3,0}$ is central, and squares to $-1$. This fact is utilized in the standard approach above to show that $\Pin{3,0}$ is isomorphic to the matrix group $\SU[\pm]{2}$. I.e. from an understanding of $\spin{3}$ and its generators, we can build up all of $\Pin{3}$ by extending via multiplication by $i$. However, $\Spin{3}$ has a quaternionic fundamental representation, as such there is an anti-linear operator which by all accounts can fill the same roll as $i$. In particular the operator:
 \begin{equation}
     J = \begin{pmatrix}
         0 & -1\\
         1 & 0 
     \end{pmatrix} *
 \end{equation}
where $*$ is complex conjugation acting to the right, commutes with every element of $\Spin{3}$ and squares to $-1$. From this we can build an exactly analogous co-representation of $\Pin{3,0}$, the elements of the algebra given by
\begin{table}[h]
    \centering
    \begin{tabular}{cll}
    $\mathbb{I}_2$ & Scalar & \\
    $s_{yz}J, \> s_{zx}J,\>s_{xy}J$  & Vector & Generates pin(3,0)\\
    $s_{yz}, \> s_{zx},\>s_{xy} $ & Bi-vector & Generates spin(3) \\
    $J$ & Pseudo-scalar & 
    \end{tabular}
\end{table}

\noindent In this case the primary reflection $\mathcal{P}$ would be represented by the anti-unitary operation of $J$ upon spin states. This would be a poor model for physics however, as the operator $J$ has no eigen-states, meaning no spin states are the same as their mirror image. Though there may be something interesting in exploring this point of view for a 3D Euclidean field theory combining these two approaches to also have access to something like a time reversal operator. 



\subsection{\texorpdfstring{Co-representation of $\spin{3,1}$}{Co-representation of Spin(3,1)}}
\label{app:worse}
The introduction of co-representations seemed most well-suited for discrete extensions of a group, and as such we have yet to consider a co-representation of an algebra. We will invent one now, in an effort to remedy the incorrect assignment of (anti-)linearity to the $\mathbf{P}$ and $\mathbf{PT}$ found from squishing the $\Pin{3,1}$ group from the real to complex case in Sec.~(\ref{sec:majasweyl}) above.

\noindent Starting with the representation of $\su{2}$ given above by $s_x, \>s_y, \> s_z$ obeying $\left[s_i, s_j\right] = \varepsilon_{ijk}\> s_k$, we build a `$\star-$complexified' algebra of operators spanned now by $s_i$ and $\K_i= s_i J$, with  $J=\varepsilon*$. The $s_i$ of course will play the same role of the rotations, but let us make sure the elements $\K_i$ could plausibly play the role of boosts. Firstly we may check the commutation relations hold, recalling that $J$ commutes with any $s_i$:
\begin{equation}
\label{eq:AntiLinComms}
\begin{split}
\left[L_i, \K_j\right] &= \left[s_i, s_j J\right] = s_i \left(s_j J\right) - \left(s_j J\right) s_i = s_i s_j J - s_j s_i J = \left[s_i, s_j\right]J = \varepsilon_{ijk} \> s_k J = \varepsilon_{ijk} \>\K_k, \\
\left[\K_i, \K_j\right] &= \left[s_i J, s_j J\right] = \left(s_i J\right) \left(s_j J\right) - \left(s_j J\right) \left(s_i J\right) = (s_i s_j - s_j s_i) J^2 =-\left[s_i, s_j\right]= -\varepsilon_{ijk} \> L_k.
\end{split}
\end{equation}
So this is in fact a realization of $\spin{3,1}$. But just because the algebra makes sense, does not necessarily mean there is some clean mapping to a group structure. Let us see the exponential map still works as we want. A basis for this co-representation gives boosts which look like:
\begin{equation}
    \K_x = s_x J=\frac{1}{2}\begin{pmatrix}
        -i & 0 \\
        0 & i
    \end{pmatrix}*\>, \quad \K_y  = s_y J=\frac{1}{2} \begin{pmatrix}
        -1 & 0 \\
        0 & -1
    \end{pmatrix}*\>, \quad \kappa_z = s_z J = \frac{1}{2}\begin{pmatrix}
        0 & i \\
        i & 0
    \end{pmatrix}*\> .
\end{equation}
For sake of clarity, define $\K_i = k_i *\>$, i.e. $k$ is the linear piece of $\K$. Then we can see explicitly
\begin{equation}
    \K_i \K_i = k_i k_i^* = \frac{1}{4}\mathbb{I}_2
\end{equation}
and so the normalization is just as before, meaning the anti-linear boosts still appear to act as non-compact generators. With this in hand consider the exponential map, at least as a formal power series:
\begin{equation}
    \exp( \beta\> \K_i) = \mathbb{I}_2+ \beta \K_i+ \frac{1}{2!}\left(\beta\kappa_i\right)^2+\ldots
\end{equation}
It is easy to see every even power of $\kappa_i$ is proportional to the identity, so every odd power will be $\kappa_i$ once again, and since all the coefficients are real, we will simply find two sums:
\begin{equation}
\begin{split}
\label{eq:anti-boost}
    \exp(\beta\> \K_i) &= \sum_{n=0}^{\infty} \frac{\left(\beta/2\right)^{2n}}{(2n)!}\mathbb{I}_2 + \sum_{i=0}^\infty\frac{\left(\beta / 2\right)^{2n+1}}{(2n+1)!}  \K_i\\
    &= \cosh\left(\beta /2\right)\mathbb{I}_2  + \sinh\left(\beta /2\right) \K_i.
\end{split}
\end{equation}
If we accept that the boosts act with a linear part and an anti-linear part, there seems to be no trouble building the action of this group. Thus we appear to have acquired a co-representation of the group $\Spin[+]{3,1}$, which we will refer to as the set of operators $\SL[\star]{2,\C}$.

\noindent Just as we did above, if we wish to build the full spin group, i.e. those elements which are pre-images of the $PT$ component of $\grp{SO}(3,1)$, then we need to find some central extension which does this. Previously the scalar $i$ played this role, however $i$ will no longer commute with every element of the group because of the anti-linearity of the boosts. Once again we can notice essentially $i$ and $J$ have switched places, and a central extension by $J$ gives us access to an entirely analogous construction, i.e. taking $J$ itself as a choice for the operator $\mathbf{PT}$.

With the full spin group accounted for, in analogy to the first construction, we may ask what another extension by $i$ will do. In particular upon the co-representation of the Lie algebra, multiplication by $i$ will commute with rotations, but of course anti-commute with the boosts $\K_i$ via the complex conjugation. This can also be seen from the expression Eq.~(\ref{eq:anti-boost}), moving $i$ past the boost will flip the anti-linear term, which has the same effect of sending $\beta \mapsto -\beta$. Thus multiplication by $i$ will perform the outer automorphism of the Lie algebra, which physically must be interpreted as a parity or time reversal. All in all one will find four sets of operators for the four components of the Lorentz group: $1, PT, P, T$
\begin{equation}
    \SL[\star]{2,\C}, \quad \SL[\star]{2,\C}J, \quad i\SL[\star]{2,\C}, \quad i\SL[\star]{2,\C}J
\end{equation}
In particular we find what we are after: if $i$ is taken to be the parity operator $\mathbf{P}$, then both $\mathbf{PT}$ and $\mathbf{T}$ are correctly anti-linear. It can be shown that this co-representation preserves the inner products:
\begin{equation}
\begin{split}
    &\varphi^\dagger \psi \pm \varphi^\intercal \psi^* \\
    &\varphi^\intercal \varepsilon\> \psi \pm \varphi^\dagger \varepsilon \>\psi^*
\end{split}
\end{equation}
Where $-$ is for when $\psi$ and $\varphi$ are from the same Weyl spinor representation, and $+$ for when they come from opposite representations. At the cost of some rather strange boosts, the Parity and Time reversal operators have the correct linearity and anti-linearity, respectively.

\newpage
\bibliographystyle{plain} 
\bibliography{Entire} 

\end{changemargin}
\end{document}